\documentclass[%
 reprint,
 amsmath,amssymb,
 aps, prd,preprintnumbers, superscriptaddress, nofootinbib
]{revtex4-2}

\usepackage{graphicx}	
\usepackage{amsmath}	
\usepackage{amssymb}	

\usepackage[english]{babel}
\usepackage{graphicx}
\usepackage{latexsym}
\usepackage{floatflt}
\usepackage{float}
\usepackage{units}
\usepackage{hyperref}
\usepackage{amsmath}
\usepackage{gensymb}
\usepackage{mathdots}
\usepackage{color}
\usepackage{xcolor}
\usepackage{mathtools}
\usepackage{lipsum}

\usepackage{dcolumn}
\usepackage{bm}

\newcommand{\id}{{\rm d}}

\begin{document}

\preprint{MIT-CTP/5892}

\title{Not-quite-primordial black holes seeded by cosmic string loops}
\author{Bryce Cyr}
\email{brycecyr@mit.edu}
\affiliation{Center for Theoretical Physics - A Leinweber Institute, Massachusetts Institute of Technology, Cambridge, MA 02139,USA}%

\begin{abstract}
Cosmic strings appear in many well-motivated extensions to the standard model of particle physics. If they exist, an abundant population of compact objects known as cosmic string loops permeate the Universe at all times, providing a secondary source of density perturbations that are large amplitude and non-gaussian in nature. In general, these loops are not stationary in the rest frame of the dark matter, thus their relative velocities will typically seed both spherical and filamentary overdensities in the matter era. Building upon previous work, we provide an improved framework to compute the complete halo mass function for these string seeded overdensities, valid for any loop velocity distribution. Using this mass function, we also compute the subset of halos capable of undergoing a direct collapse, forming a population of black holes with initial mass $10^{4-5} \, M_{\odot}$ at high redshifts. Interestingly, for reasonable values of the string parameters, one can reproduce the abundance of ``Little Red Dots" as inferred by JWST.
\end{abstract}

\maketitle

\section{\label{sec:level1}Introduction}
Over the past two decades, observations made by the Hubble Space Telescope (HST), James Webb Space Telescope (JWST), and numerous other instruments, have uncovered an abundant population of massive objects at high redshifts \cite{Bowler2020,Harikane2021,Bouwens2021,Harikane2023,Leung2023,Perez2023,Harikane2024,Adams2023,Willot2024,Finkelstein2024,Donnan2024,Whitler2025, Matteri2025}. Most recently, the unparalleled sensitivity of JWST has allowed us to discover a previously undetected sample of red, compact galaxies dubbed \textit{little red dots} \cite{Labbe2022,Kocevski2023, Matthee2023, Pacucci2023, Durodola2024} which are also thought to host supermassive black holes. Combined with observations of high redshift quasars, the Universe appears to be much more active at early times than was previously thought.

The presence of supermassive black holes at high redshifts \cite{Maiolino2023} is particularly troubling from the perspective of usual galaxy formation models. Assuming a standard $\Lambda$CDM Universe, the first black holes form from the deaths of Population III stars. In principle, this could lead to a distribution of black holes with initial mass $M_{\rm BH} \simeq 100 \, {M_{\odot}}$ at $z \simeq 20$. However, even under the optimistic assumption of continuous Eddington accretion from formation to detection (at e.g. $z \simeq 7-10$), these so-called light seeds \cite{Volonteri2005,Volonteri2010, Volonteri2021, Natarajan2011} typically fall more than an order of magnitude below the masses indicated by current observations\footnote{Some notable examples include GN-z11 \cite{Bunker2023} ($M_{\rm BH} \simeq 1.6 \times 10^6 \, M_{\odot}$, $z \simeq 10.6$), CEERS 1019 \cite{Larson2023} ($M_{\rm BH} \simeq 10^7 \, M_{\odot}$, $z \simeq 8.7$), and UHZ1 \cite{Natarajan2023} ($M_{\rm BH} \simeq 4 \times 10^7 \, M_{\odot}$, $z \simeq 10.1$).}. 

With the light seed origin of supermassive black holes facing significant tension at high redshifts, more attention has been placed on formation mechanisms for ``heavy seeds", in which black holes are created with an initial mass in the $10^{4-6} \, M_{\odot}$ range. With heavy seeds, a more modest accretion rate and merger history may be used to grow these black holes in a way that allows them to match observations \cite{Natarajan2011,Bogdan2023, Natarajan2023}. One such mechanism capable of producing heavy seeds is through runaway mergers of Pop III stars in the first galaxies \cite{Portegies2002, Boekholt2018}, though this typically requires extremely dense environments. Another mechanism is through the monolithic collapse of metal (and molecular hydrogen)-free gas clouds in the early Universe \cite{Haehnelt1993,Umemura1993, Loeb1994, Eisenstein1994,Haiman1999, Oh2001, Bromm2002, Begelman2006, Inayoshi2019, Begelman2025}, which are thought to form objects known as direct collapse black holes (DCBHs). Satisfying the DCBH conditions within standard $\Lambda$CDM is in general difficult due to the abundance of molecular hydrogen (${\rm H}_2$) in the first star forming halos at $z\simeq 20$.

It has recently been shown, however, that enhanced primordial density perturbations on small scales ($k \lesssim 1 \,\,{\rm Mpc}^{-1}$) can lead to the early collapse and subsequent virialization of massive halos at high redshifts ($z \gtrsim 200$) \cite{Qin2025}. At these redshifts, the cosmic microwave background (CMB) is sufficiently energetic to inhibit the formation of ${\rm H}_2$, thus providing an ideal environment for the production of direct collapse black holes. The authors of \cite{Qin2025} introduce the nomenclature \textit{not-quite-primordial black holes} (NQPBH) to describe the formation of DCBHs at high redshift, which we adopt here. Their results are broadly consistent with other literature on early structure formation from enhanced small scale perturbations \cite{Hirano2015,Ito2024}.

The most robust constraint on the power spectrum at these small scales comes from the production of CMB spectral distortions\footnote{Other constraints, coming from heating of ultrafaint dwarfs \cite{Graham2024}, early reionization, and CMB accretion \cite{Bringmann2025} may also be relevant depending on the seed scenario.}, which are generated through the silk damping of these enhanced perturbations \cite{Chluba2012, Cyr2023b} occurring at $10^3 \lesssim z \lesssim 10^6$. As an integrated constraint, spectral distortions generated by the mild step function considered in \cite{Qin2025} ($P_{\zeta}(k>10^2 \, {\rm Mpc}^{-1})\simeq 10^{-7}$) are already at the upper bound provided by the COBE/FIRAS experiment.

In light of this, it is prudent to revisit other scenarios which are capable of forming sufficiently massive halos at early times. One such scenario is that of the formation of structure around a distribution of cosmic string loops. Cosmic strings can form during cosmological phase transitions in the very early Universe, and appear in a wide variety of well motivated extensions to the standard model of particle physics. Heuristically, one can think of a cosmic string loop as a highly localized region of space with a very high mass density, similar in spirit to a primordial black hole (PBH). From the perspective of structure formation, however, cosmic strings differ from PBHs in two major ways. First, string loops are sourced continuously at all redshifts, generically leading to a very wide mass distribution. Second, loops are typically produced with a very high velocity relative to the dark matter, with numerical simulations indicating a central value for the velocity distribution of $\langle v_{\rm form} \rangle \simeq 0.3$. 

More recently, early structure formation around cosmic string loops has seen a renewed interest \cite{Shlaer2012,Jiao2023,Jiao2024,Jiao2023b, Koehler2024} in light of challenges to current astrophysical paradigms stemming from the aforementioned observations. With some notable exceptions \cite{Olum2006,Shlaer2012}, much of this work has centered around accretion onto stationary objects due to the simplicity of the growth rate. While a valuable first step, we will show that only a very small fraction of loops are expected to accrete in this stationary spherical limit. Thus, we improve this structure formation picture by combining and extending the formalisms set out by \cite{Shlaer2012} and \cite{Jiao2023} to the case of a non-trivial distribution of loop velocities. Importantly, the formation of string loops do not produce large acoustic waves, thus they are unconstrained by CMB spectral distortion bounds \cite{Tashiro2012}.

The main purpose of this work is twofold. First, we wish to provide a self-consistent framework which describes the formation and evolution of dark matter overdensities and halos around string loops with any string tension, and at any loop velocity. Second, under reasonable choices for these string parameters, we wish to compute the abundance of halos which are capable of seeding direct collapse black holes. As we will show, black hole abundances produced in these string-seeded scenarios can be consistent with those inferred from observations of high redshift quasars and little red dots. 

The structure of this paper is as follows. In Sec.~\ref{sec:Cosmic-string}, we provide a basic introduction to the theory of cosmic strings, including the expected abundance of loops and general constraints on the string tension. Sec.~\ref{Sec:spherical_growth} provides an overview of the accretion rate in the case of spherical, stationary loops. The growth rate around moving loops is reviewed (and improved) following \cite{Shlaer2012} in Sec.~\ref{sec:cylindrical_growth}. We provide a unified framework and derive the mass function of string seeded overdensities in Sec.~\ref{sec:direct_collapse}. Particular emphasis is put on determining the subset of these overdensities which can host direct collapse black holes, which we discuss in detail, before commenting on the expected black hole abundances. We discuss a variety of future directions and conclude in Sec.~\ref{sec:discussion}. Throughout, we use natural units in which $\hbar = c = k_{\rm b} = 1$.

\section{Cosmic strings} \label{sec:Cosmic-string}
Cosmic strings are one dimensional topological defects that can form when the Universe undergoes a phase transition \cite{Vilenkin2000, CS1,CS2}. The condition for the formation of cosmic strings is that the true vacuum manifold ($\mathcal{M}$) of the field undergoing the phase transition must not be simply connected. The prototypical case for this is when the true vacuum manifold is that of a circle, $\mathcal{M} = S^1$. If such a phase transition takes place, the Kibble mechanism \cite{Kibble1, Kibble2} dictates that the formation of cosmic strings is unavoidable in our expanding Universe, and a network of such defects is established. The mass per unit length ($\mu$) of the string is related to the scale of the phase transition $(\eta)$ by the proportionality $\mu \simeq \eta^2$, and constraints are usually set on the dimensionless quantity $G\mu$, where $G$ is Newton's gravitational constant.

Immediately after formation, the network exists as a tangled web, which relaxes quickly to a scaling regime in which the initial conditions of the phase transition are washed out\footnote{Simulations \cite{Albrecht1989,Allen1990,Bennett1989,Martins1996,Vincent1997} indicate that it takes roughly $\mathcal{O}(10)$ Hubble times from string formation to the scaling regime, which at very early times is incredibly rapid.}. At this time, a few ($N_{\ell}$) long strings exist, running through the entire Hubble patch, with total energy density given by $\rho_{\infty} \simeq N_{\ell} \, \mu \,H^2$. Consequently, their fractional contribution to the energy density of the Universe remains frozen in, as $\Omega_{\rm CS} = \rho_{\infty}/\rho_{\rm crit} \simeq N_{\ell} \, G\mu$. One way of searching for these long strings is by looking for jumps in the CMB temperature anisotropies (the so-called Gott-Kaiser-Stebbins effect \cite{Kaiser1984,Gott1984}) which are induced due to the conical spacetime around a string. A non-detection of this effect in the Planck/WMAP \cite{Wyman2005,Bevis2007,Planck2013} data therefore puts a robust bound on $G\mu$ at the level of $G\mu \lesssim 10^{-7}$.

As these long strings evolve, they will experience intersections and self-intersections, sourcing a distribution of cosmic string loops, which also appear to follow a scaling solution \cite{Albrecht1989,CSsimuls2,CSsimuls3,CSsimuls4,CSsimuls5,CSsimuls6,CSsimuls7,CSsimuls8,CSsimuls9,CSsimuls10}. The loops themselves are typically formed with an initial size $L_{\rm form} \simeq \alpha t_{\rm form}$ where $\alpha \simeq 0.1$ is inferred from simulations. After being produced, an individual loop will oscillate under its own tension, radiating gravitational waves and slowly evaporating. The energy loss in gravitational waves is a well studied consequence of these oscillations, and is given by
%
\begin{align}
    P_{\rm g} \simeq \Gamma_{\rm g} G\mu^2,
\end{align}
%
where $\Gamma_{\rm g} \approx 50$ \cite{Vachaspati1984}. Assuming this to be the only source of energy loss, the size of a string loop will evolve as 
%
\begin{align}
    L(t) = L_{\rm form} - \Gamma_{\rm g} G\mu(t-t_{\rm form}).
\end{align}
%
Generally, $\alpha \gg \Gamma_{\rm g} G\mu$ leading to string loops that are quite long-lived. Their decay and formation times can be related through $t_{\rm dec} \simeq (\alpha/\Gamma_{\rm g} G\mu) t_{\rm form}$. Sophisticated simulations \cite{CSsimuls8} of loop production in the Nambu-Goto limit (neglecting the intrinsic thickness of the string) have provided us with the spectrum of these loops in the radiation and matter eras
%
\begin{align}
    \label{eq:dNdLradiation}
    \left.\frac{\id N}{\id L}\right|_{\rm r} &= \frac{\alpha_{\rm r}}{t^{3/2}(L+\Gamma_{\rm g} G\mu t)^{5/2}} \times \begin{cases} 1 &(t \leq t_{\textrm{eq}})
    \\
    \left(\frac{t_{\textrm{eq}}}{t}\right)^{1/2} &(t>t_{\textrm{eq}}),
    \end{cases}\\
    \left.\frac{\id N}{\id L}\right|_{\rm m} &= \frac{\alpha_{\rm m}}{t^{2}(L+\Gamma_{\rm g} G\mu t)^{2}}, \label{eq:dNdLmatter}
\end{align}
%
where $N$ is the physical number density of loops. The first line corresponds to loops that are produced in the radiation era, which may persist into the matter era before evaporating, while the second line is specifically for loops produced after matter-radiation equality ($t > t_{\rm eq}$). The normalization coefficients are extracted from simulation and found to be $\mathcal{O}(0.1)$ \cite{CSsimuls8}. We will use values of $\alpha_{\rm r} = 0.18$, and $\alpha_{\rm m} = \alpha_{\rm r}/\sqrt{\alpha} \approx 0.57$ \cite{Cyr2023} which ensures that the spectrum remains piecewise continuous. Heuristically, at a given time $t$, the spectrum of loops grows with a power law behaviour until a critical size, $L_{\rm crit} \simeq \Gamma_{\rm g} G\mu t$, below which the distribution flattens off. Loops with $L<L_{\rm crit}$ decay within one Hubble time, while $L>L_{\rm crit}$ are referred to as ``long-lived". The gravitational radiation produced by a distribution of loops is known to source a stochastic background. Under the assumption that the background detected by the pulsar timing array consortium does not originate from loops, one can set a stronger bound \cite{NANOGravDetection2023, EPTADetection2023, PPTADetection2023, CPTADetection2023, NANOGrav2023Exotic} on the string tension of $G\mu \lesssim 10^{-10}$. This bound is thought to be less robust than the CMB limit, as it depends on the more uncertain dynamics of the loop distribution, thus we will mainly be concerned with string tensions satisfying $G\mu \lesssim 10^{-7}$. For example, recent work \cite{Zhao2025} has shown that gravitational wave emission from local strings is suppressed relative to direct particle production, while for global strings the Nambu-Goto approximation may break down as the core of the string becomes delocalized. 

Cosmic strings are solitonic objects, meaning that they represent a region of trapped energy density. This energy density gravitates, thus over time, cosmic string loops will acquire dark matter halos. In the past, this fact has been used to study a variety of effects, such as the possibility of early reionization \cite{Olum2006}, additional structures at early times \cite{Shlaer2012,Jiao2023b}, and the ability of string-seeded halos to host intermediate mass black holes \cite{Brandenberger2021}. More recently, N-body \cite{Jiao2024} and hydrodynamic \cite{Koehler2024} simulations have been employed to study the evolution of structure around a (stationary) cosmic string loop. Those articles seem to indicate that the inclusion of these objects can alleviate tensions between the abundant population of high redshift UV bright galaxies detected by the James Webb space telescope (JWST), and the standard galaxy formation scenario assuming only Gaussian density perturbations.

In addition to this numerical work, there have been advancements in our analytic and semi-analytic understanding of the precise growth rate experienced by a string-seeded dark matter halo \cite{Jiao2023}. These improvements come mainly from treating the string as an extended object instead of a point mass when considering the turn-around surface of a given shell of gas/dark matter. We further extend these improvements by considering the effects of loop velocity in what follows. 

Recently, Qin \textit{et al.} have shown that sufficiently massive overdensities present at high redshift ($z \gtrsim 200$) yield near-optimal conditions for direct-collapse black hole formation. We note that DCBH formation in the context of superconducting cosmic strings was considered in \cite{Cyr2022}, where the direct collapse conditions can be satisfied at low redshifts due to an external photon flux coming from the strings themselves. 

In the following subsections, we show how cosmic string loops can provide a viable source of these early massive halos, without introducing an enhancement to the primordial power spectrum. This allows us to fully avoid bounds set by spectral distortions to the cosmic microwave background \cite{Chluba2012, Cyr2023b}, freeing up some regions of parameter space that would otherwise have been constrained. Following that, we discuss in more detail how the direct collapse conditions are satisfied in the cosmic string case, before finally computing the expected abundance in Sec.~\ref{sec:direct_collapse}.

\subsection{Spherical growth around a stationary loop} \label{Sec:spherical_growth}

%
\begin{figure*}
\centering 
\includegraphics[angle = 270, width=1.9\columnwidth]{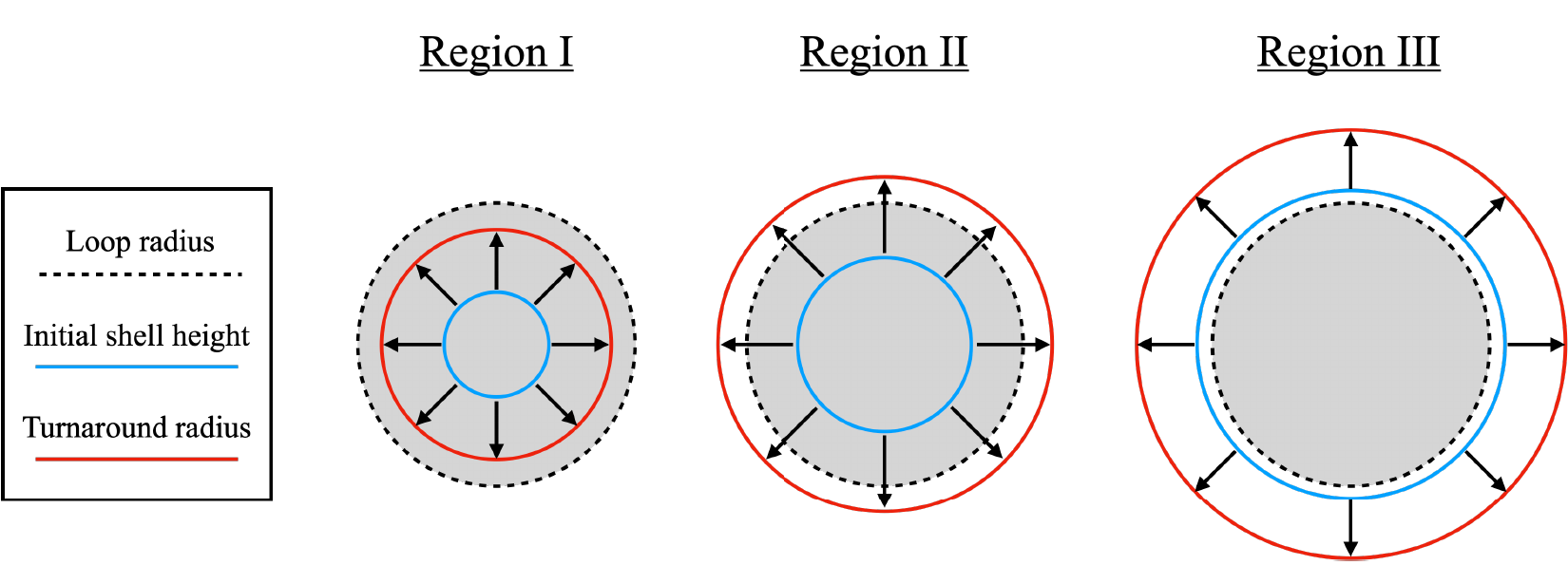}
\caption{A heuristic description of the three different phases of accretion experienced by the string-seeded overdensity, in the stationary limit. The point mass approximation is strictly valid in Region III. Figure adapted from \cite{Jiao2023}.}
\label{fig:Growth-phases}
\end{figure*}
%

The accumulation of dark matter around an oscillating cosmic string loop was considered in detail in Hao \textit{et al.} \cite{Jiao2023} which we now summarize. Consider a string loop comoving with the Hubble flow at some time $t$, whose centre of mass sits at the origin of a polar coordinate system with radial direction $r$. The standard picture is that spherical shells surrounding the string loop will decouple from the Hubble flow and fall into the overdensity, slowly building up a halo of dark matter. 

There are two important scales to keep in mind. First, the turnaround radius $r_{\rm TA}$ is the physical scale at which a given spherical shell is stationary at specific point in time, i.e. the moment in which the shell has ceased to expand with the Hubble flow. After this point, the shell begins its collapse into the string-seeded halo. Second, the virial radius ($r_{\rm vir}$) is the scale at which shell-crossings occur, leading to the homogenization of the halo and establishing a virial core. The virial radius is usually taken to be a reasonable proxy for the size of a given halo. These two scales are roughly related by $r_{\rm TA} \simeq 2 r_{\rm vir}$, and the Zel'dovich approximation can be used to compute $r_{\rm TA}$ in the context of cosmic string loops \cite{Zel'dovich,Jiao2023}.

Imagine now that we have a cosmic string loop with total length $L$. Its approximate radius of curvature is given by $R = L/\sigma$, where $\sigma = 2\pi$ would correspond to a perfectly circular loop. The loop itself undergoes relativistic oscillations about its centre of mass with period $T\simeq 2\pi R$, which allows us to derive a time-averaged density profile \cite{Jiao2023}
%
\begin{align} \label{eq:smeared-profile}
    \bar{\rho}(r) = \frac{\sigma \mu}{2 \pi^2 r^2 \sqrt{1-(r/R)^2}}.
\end{align}
%
This expression is valid provided that the oscillation timescale of a loop is rapid compared to the typical turnaround time of a mass shell. As a result, we will primarily be interested in understanding the growth rate of overdensities forming around cosmic string loops which are sourced in the radiation epoch, but persist into the matter dominated era. Quantitatively, this means that we will be considering loops forming at $z_{\rm form} \geq 2 z_{\rm eq}$, roughly a Hubble time before matter radiation equality. Specifically, $\bar{\rho}(r)$ describes how the mass distribution of the oscillating loop is smeared out over its maximal radial extent, from $0 < r \leq R$. 

The cosmic string loop is a bound object, whose physical extent is fixed by initial conditions at $R$, whereas spherical shells of matter will initially expand (in physical coordinates) with the Hubble flow before falling into the overdensity. Qualitatively, this means that the gravitational attraction felt by a given spherical shell will depend on the initial position of the shell relative to the loop radius $R$. In Figure~\ref{fig:Growth-phases}, we show the three different scenarios that can occur. 
\begin{itemize}
    \item \textit{Region I}: The spherical shell starts ($r_{\rm i}$) and turns around ($r_{\rm TA}$) inside the extent of a string loop. Namely, $r_{\rm i} < r_{\rm TA} < R$.
    \item \textit{Region II}: The spherical shell starts inside the string loop, but turns around after radius-crossing. Here, $r_{\rm i} < R < r_{\rm TA}$.
    \item \textit{Region III}: The initial size of the spherical shell is larger than the loop, which implies $R < r_{\rm i} < r_{\rm TA}$.
\end{itemize}
The point-mass approximation\footnote{This approximation assumes the entire loop mass is located at $r = 0$, in other words, that $\bar{\rho}(r) = \frac{\sigma R \mu}{4\pi r^2} \, \delta(r)$.} for the loop is strictly speaking only valid in Region III, when a spherical shell always feels the full extent of the gravitational potential sourced by the string. Regions I and II represent important corrections to the growth rate at the early stages of halo formation.

Accretion onto a compact object can be well-described using the Zel'dovich approximation\footnote{A more sophisticated analysis gives an $\mathcal{O}(1)$ correction to the accretion process \cite{Fillmore1984,Bertschinger1985,Bertschinger1987}.}, which aims to model the evolution of matter shells around the cosmic string loop. Consider a comoving scale $q$, defined relative to the centre of the loop. The physical distance from the loop centre is given by
%
\begin{align}
    h(q,z) = a[q - \psi(q,z)],
\end{align}
%
where $a$ is the scale factor (normalized such that $a(t_0) = 1$), and $\psi$ is the perturbation induced on the comoving shell due to the compact object. In the absence of a string loop, the shell would simply expand with the Hubble flow. A shell turns around and begins to collapse onto the overdensity when $\id h(q_{\rm nl},\id_{\rm nl})/dz = 0$. One can show that this condition implies
%
\begin{align}
    q_{\rm nl} = -(1+z)^2 \frac{\id}{\id z} \left[\frac{\psi(q_{\rm nl},z)}{1+z} \right].
\end{align}
%
Thus, if one solves for the evolution of the perturbation, $\psi(q_{\rm nl},z)$, it is possible to determine the size of a shell that turns around at any given redshift, $z$. The general evolution equation for $\psi$ is
%
\begin{align} \label{eq:pert-eq-gen}
    \frac{\id^2 \psi}{\id z^2} + \frac{1}{2(1+z)} \frac{\id \psi}{\id z} - \frac{3}{2(1+z)^2} \psi = \frac{9G M(aq)t_0^2}{4q^2(1+z)^2}. 
\end{align}
%
The derivation of this expression is detailed in both \cite{Vilenkin2000}, as well as Appendix A of \cite{Jiao2023}. Conservatively, we neglect growth in the radiation era and use initial conditions $\psi(z_{\rm eq}) = \id \psi(z_{\rm eq})/\id z = 0$. The mass (of the loop) enclosed within a comoving shell is given by $M(aq)$, determined through 
%
\begin{align}
    M(aq) = \int_0^{aq} \id r \, 4 \pi r^2 \bar{\rho}(r),
\end{align}
%
where $\bar{\rho}$ is the smeared density profile given in Eq.~\eqref{eq:smeared-profile}. Once $q_{\rm nl}$ has been found, the associated mass of the halo is simply given by
%
\begin{align}
    M_{\rm nl}(z) = \frac{4\pi}{3} \rho_{\rm bg} q_{\rm nl}^3(z).
\end{align}
%
where $\rho_{\rm bg} = 3 H_0^2/8\pi G$ is the comoving critical background density. \\

\subsubsection{Point mass limit (Region III growth)}
In this limit, the enclosed mass simply becomes $M(aq) = \mu L$, and Eq.~\ref{eq:pert-eq-gen} has a fully analytic solution,
%
\begin{align}
    \psi^{\rm III}(q,z) &= \frac{9}{10}\frac{G \mu L t_0^2}{q^2}\left(\frac{1+z_{\rm eq}}{1+z}\right) \nonumber\\
    &\times \left[1 - \frac{5}{3}\left( \frac{1+z}{1+z_{\rm eq}}\right) + \frac{2}{3}\left( \frac{1+z}{1+z_{\rm eq}}\right)^{5/2} \right]. \nonumber
\end{align}
%
In this case, we find that the scale turning around at redshift $z$ is given by
%
\begin{align} 
    q_{\rm nl}^{\rm III}(z) &= \bigg(\frac{9}{5} G \mu L t_0^2 \left(\frac{1+z_{\rm eq}}{1+z}\right) \nonumber\\
    &\times \left[1 - \frac{5}{6}\left(\frac{1+z}{1+z_{\rm eq}} \right) - \frac{1}{6}\left(\frac{1+z}{1+z_{\rm eq}} \right)^{5/2} \right] \bigg)^{1/3}. \nonumber
\end{align}
%
From here, it is straightforward to determine the total halo mass at any point,
%
\begin{align} \label{eq:Mnl_III}
    M^{\rm III}_{\rm nl}(z) = \frac{2}{5} \mu L \left(\frac{1+z_{\rm eq}}{1+z}\right) \tilde{M}(z),
\end{align}
%
where we have approximated $H_0 = 2/(3t_0)$. The term $\tilde{M}(z)$ captures effects relevant close to $z_{\rm eq}$, and has the asymptotic limit $\tilde{M}(z\ll z_{\rm eq}) \simeq 1$. Its exact form can be found in Appendix \ref{sec:appendix_a}. Shortly after matter radiation equality, the overdensity exhibits a growth-rate linear in the scale factor, $M_{\rm nl}(z) \simeq M_{\rm loop} (a/a_{\rm eq})$. It is important to note that the point mass approximation is valid for all shells with $q > R(1+z_{\rm eq})$, as they originate from beyond the string loop at matter-radiation equality. Recalling that the loop radius is related to the formation time by $R \simeq (\alpha/\sigma)t_{\rm form}$, we can use this fact to establish a relationship between the redshift of formation of a loop ($z_{\rm form}$), and the redshift at which the point mass approximation is valid ($z_{\rm III}$). 

\begin{figure}
\includegraphics[width=0.95\columnwidth]{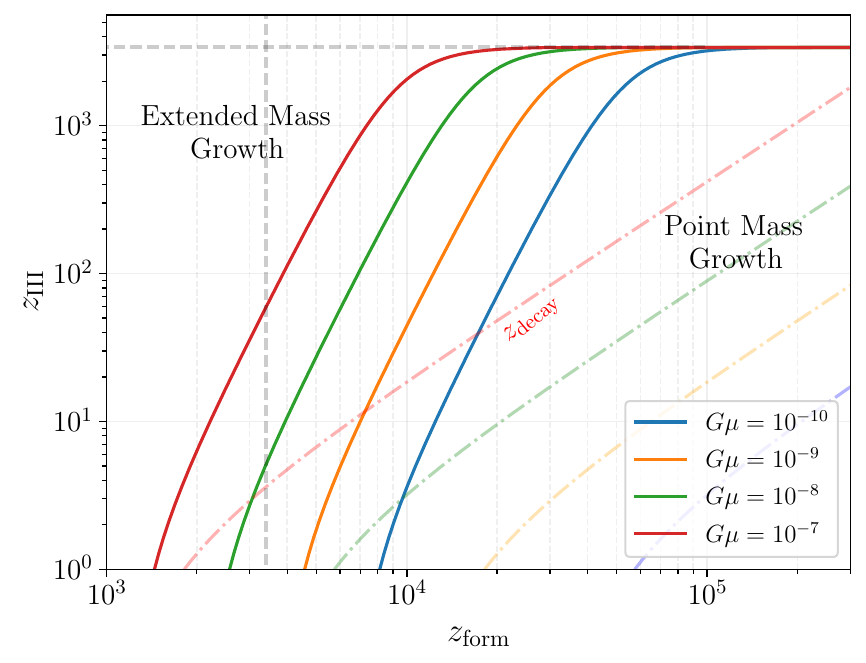}
\caption{Contours showing the redshift at which the point mass approximation is valid, for a given formation redshift $z_{\rm form}$ and string tension $G\mu$. Above each contour, turnaround shells originate inside the loop, while below the point mass approximation given in Eq.~\eqref{eq:Mnl_III} is valid. We also plot the redshift that a given loop decays using the dash-dotted line, and dashed lines indicate $z_{\rm eq}.$}
\label{fig:point_mass_validity}
\end{figure}

This relation is shown in Fig.~\ref{fig:point_mass_validity} for different values of the string tension, $G\mu$. For each string tension, the point mass limit is valid beneath its respective curve. For lower values of the string tension, shells take longer to turn around and the point mass limit is only applicable for very small loops. The dashed horizontal and vertical lines correspond to $z_{\rm eq}$. The smeared density profile $\bar{\rho}$ is not valid for loops forming after matter radiation equality as the typical oscillation timescale is close to a Hubble time. In this work, we have implicitly assumed that the size of a given loop does not shrink substantially before the halo achieves a mass of $M_{\rm nl} \simeq M_{\rm loop}$. After this point, the overdensity becomes self-sustaining and will continue to grow even if the loop disappears. For a chosen value of $G\mu$, this assumption is justified at all points above the $z_{\rm decay}$ contour on Fig.~\ref{fig:point_mass_validity}. This will always be the case for loops forming at $z_{\rm form} \lesssim {\rm few} \times 10^{5}$ for $G\mu \lesssim 10^{-7}$.

%
%

\subsubsection{Collapse of interior shells (Region I/II growth)}
Shells that begin their journey at $z_{\rm eq}$ with $q < R(1+z_{\rm eq})$ cannot be described by the point mass approximation, as they only feel the gravitational attraction from a fraction of the total loop mass. Before crossing the loop radius, the enclosed mass has another form, given by
%
\begin{align}
    M(aq) = \frac{2}{\pi} \mu \sigma R \arcsin\left(\frac{aq}{R} \right).
\end{align}
%
For shells that originate and turn around near the loop centre, one can approximate $\arcsin(aq/R) \simeq aq/R$. In this limit, the non-linear mass again has a simple analytic form, given by
%
\begin{align}
    M^{\rm I}_{\rm nl}(z) &= \frac{2 t_0}{9G}\bigg(\frac{18\sigma}{5\pi} \frac{G\mu }{1+z} \bigg)^{3/2}\nonumber\\
    &\times \bigg[ \ln\left(\frac{1+z_{\rm eq}}{1+z} \right) + \frac{1}{10} - \frac{1}{10} \left(\frac{1+z}{1+z_{\rm eq}} \right)^{5/2} \bigg]^{3/2}.
\end{align}
%
As one may have expected, growth in Region I is independent of the total loop radius, $R$. From the perspective of shells in Region I, they see an ever-increasing source of energy density as they expand. 

It was found in \cite{Jiao2023} that growth in Region II follows the same redshift dependence as in Region I, thus it is reasonable to estimate $M_{\rm nl}^{\rm I} \simeq M_{\rm nl}^{\rm II}$. One can do slightly better than this by demanding that at $z_{\rm III}$ the total accreted mass matches between Regions I and III. Specifically we define $M^{\rm I/II}_{\rm nl}(z) \equiv A \, M_{\rm nl}^{\rm I}(z)$, where $M^{\rm I/II}_{\rm nl}(z_{\rm III}) = M^{\rm III}_{\rm nl}(z_{\rm III})$. Provided that $z_{\rm III} \ll z_{\rm eq}$, we find that the fudge factor $A$ can be well approximated by
%
\begin{align} \label{eq:fudge_factor_approx}
    A \simeq \left[ \left(\frac{2}{\pi}\right)\log\left( \frac{1+z_{\rm eq}}{1+z_{\rm III}}\right) \right]^{-3/2},
\end{align}
%
where $z_{\rm III}$ can be read off of Fig.~\ref{fig:point_mass_validity} for a string loop formed at redshift $z_{\rm form}$ with tension $G\mu$. In the case where $z_{\rm III} \simeq z_{\rm eq}$, one should use the full expression for the fudge factor given in Appendix~\ref{sec:appendix_a}. The accumulation of matter around a stationary loop can be reasonably approximated by
%
\begin{align} \label{eq:Mnl_all}
    M_{\rm nl}(z) = \begin{cases}
        M_{\rm nl}^{\rm I/II}(z)  \hspace{8mm} (z>z_{\rm III}),\\
        M_{\rm nl}^{\rm III}\,\,(z) \hspace{8mm} (z\leq z_{\rm III}).
    \end{cases}
\end{align}
%
This form for the total accreted mass in a halo is valid for loops moving slow relative to the dark matter rest frame, which we elaborate on in the following subsection.

\subsection{Cylindrical growth around a moving loop} \label{sec:cylindrical_growth}
The previous subsection gave us an understanding of how an extended object acquires its dark matter halo when it is stationary with respect to the Hubble flow. Simulations \cite{CSsimuls8} indicate, however, that string loops are typically produced with a non-trivial relative velocity which we denote as $v_{\rm form}$. Without invoking model-dependent couplings between the strings and the standard model sector, the only damping expected of this velocity is due to expansion. Thus, the velocity at some later time when the loop is contributing to structure formation is related to its formation velocity by $v(z) = v_{\rm form}(1+z)/(1+z_{\rm form})$.

The collapse of matter onto a moving cosmic string loop was studied in detail by Bertschinger \cite{Bertschinger1987}, who used the Zel'dovich approximation to show that the total mass of a filament is given by
%
\begin{align} \label{eq:accretion_v}
    M_{\rm fil} = \frac{3}{5} \mu L \left( \frac{1+z_{\rm eq}}{1+z}\right) \tilde{M}(z).
\end{align}
%
Interestingly, the total accreted mass was found to be independent of the velocity of the loop, and receives a small boost relative to the stationary (point mass) case of $M_{\rm fil}(z) = (3/2) M_{\rm nl}^{\rm III}(z)$. Unlike the stationary case, however, the shape of the turnaround surface is highly eccentric, resembling more of a cylindrical wake rather than a spherical overdensity. The overall length of the filament is simply given by the ballistic trajectory of the string loop, 
%
\begin{align}
    L_{\rm fil} = 3 v_{\rm form} t_{\rm eq} \left(\frac{1+z_{\rm eq}}{1+z} \right)\left(\frac{1+z_{\rm eq}}{1+z_{\rm form}} \right) \tilde{L}(z),
\end{align}
%
where we have defined $\tilde{{L}}(z)$ similar to $\tilde{{M}}(z)$ in that $\tilde{L}(z \ll z_{\rm eq}) \simeq 1$, and has the explicit form
%
\begin{align}
    \tilde{L}(z) = 1 - \left( \frac{1+z}{1+z_{\rm eq}}\right)^{1/2}. \nonumber
\end{align}
%
To go beyond this, we follow closely the formalism of \cite{Shlaer2012}, who studied the possible longitudinal and transverse instabilities that a string-seeded overdensity can experience. These instabilities can lead to collapse and in many cases fragmentation of the cylindrical overdensity, yielding many possible dark matter halos from a single string loop. 

Shlaer \textit{et al.} \cite{Shlaer2012} showed that moving loops can also undergo two different regimes of accretion. In the first regime, turnaround surfaces begin inside the loop radius, similar to our Region I accretion above. These authors simply neglect any growth of the overdensity when this is the case. Once the turnaround surface exceeds the loop radius, their formalism requires that the halo experience an ``accelerated" growth regime, before eventually catching up to the so-called ``normal" growth limit (the second regime) where the overdensity grows according to Eq.~\eqref{eq:accretion_v}.  

In our analysis we consider that the filament grows according to the ``normal" growth limit at all times after matter radiation equality. This is done for two main reasons. First, as was shown in the stationary case, even shells that begin inside the loop radius feel some fraction of the strings gravity due to relativistic oscillations, which serve to smear out the density profile. Secondly, while at some point $z_{\rm i}$, a particular turnaround shell may be inside the string loop, at some later time $z_{\rm f}$ the loop has moved relative to the shell and more of this smeared density profile is felt by the test particles. For rapidly moving loops, these two times are rather close together, thus Eq.~\eqref{eq:accretion_v} should yield a reasonable approximation at all times\footnote{Properly incorporating the effects of both the relative loop velocity and the relativistic oscillations is beyond the scope of this work, though some preliminary steps to understanding this have been presented in \cite{Jiao2023}.}.

\subsubsection{Longitudinal filament collapse}
String loops moving with a low enough relative velocity at matter radiation equality can suffer from longitudinal collapse, which, for the purposes of this work, means that the stationary spherical accretion discussed in the last subsection can be applied. To roughly estimate the necessary conditions for spherical collapse,
we can compare the contributions to the expansion of the filament coming from the Hubble flow, to the piece developed due to the peculiar motion of the loop. For a filament of length $L_{\rm fil}$, these two velocities are given by \cite{Shlaer2012}
%
\begin{align}
    v_{\rm Hubble} &\simeq \frac{L_{\rm fil}}{t_0}(1+z)^{3/2}, \nonumber\\
    v_{\rm long} &\simeq \frac{G M_{\rm fil}}{L_{\rm fil}^2} \frac{t_0}{(1+z)^{3/2}} \nonumber.
\end{align}
%
When $v_{\rm Hubble} \gtrsim v_{\rm long}$, spherical collapse occurs as the motion of the loop provides only a subdominant contribution to elongating the turnaround surface. Equating these two yields the redshift below which spherical collapse should commence
%
\begin{align} \label{eq:redshift_sc}
    (1+z_{\rm sc})\frac{\tilde{L}^3(z_{\rm sc})}{\tilde{M}(z_{\rm sc})} = \frac{\alpha}{45v_{\rm form}^3} G\mu (1+z_{\rm form}).
\end{align}
%
Recall that for $z \ll z_{\rm eq}$, we simply have $\tilde{L} \simeq \tilde{M} \simeq 1$, so we can usually neglect this contribution. As expected, the favourable conditions for spherical collapse are high string tensions, early formation times, and low formation velocities.

\subsubsection{Transverse filament collapse}
As we will see, the vast majority of parameter space yields $z_{\rm sc} \leq 0$, which means spherical collapse is never justified. In this case, one can instead consider instabilities which occur along the direction transverse to the symmetry axis of the cylindrical overdensity. These transverse instabilities will cause the cylindrical overdensity to fragment into many near-spherical overdensities, referred to in this context as \textit{beads}.

Shlaer, Vilenkin, and Loeb \cite{Shlaer2012} first considered this fragmentation process in the context of string loops. At $z \leq z_{\rm eq}$, the filament increases in longitudinal size from the motion of the loop, while also increasing in radial extent from the subsequent infall of dark matter onto the cylinder itself. These authors posited that once this radial extent exceeds the Jeans length at a particular redshift, the instability is triggered and the cylinder fragments into many near-spherical regions \cite{Quillen2010, Shlaer2012}. The physical size of the largest of these beads was determined to be
%
\begin{align}
    L_{\rm bead} = 20 \pi t_{\rm eq} \left[ \frac{\alpha G\mu}{v_{\rm form}} \left( \frac{1+z_{\rm eq}}{1+z}\right)^3 \left(\frac{1+z_{\rm eq}}{1+z_{\rm form}} \right) \right]^{1/2}.
\end{align}
%
The average number of beads expected in the fragmentation process is thus $n_{\rm bead} = L_{\rm fil}/L_{\rm bead}$, given explicitly as
%
\begin{align} \label{eq:N_bead}
    n_{\rm bead} = \frac{3 \tilde{L}(z)}{20\pi} \left(\frac{v_{\rm form}^3}{\alpha} \right)^{1/2} \left( \frac{1+z}{1+z_{\rm form}}\right)^{1/2} (G\mu)^{-1/2}.
\end{align}
%
%
\begin{figure*}
\centering 
\includegraphics[width=\columnwidth]{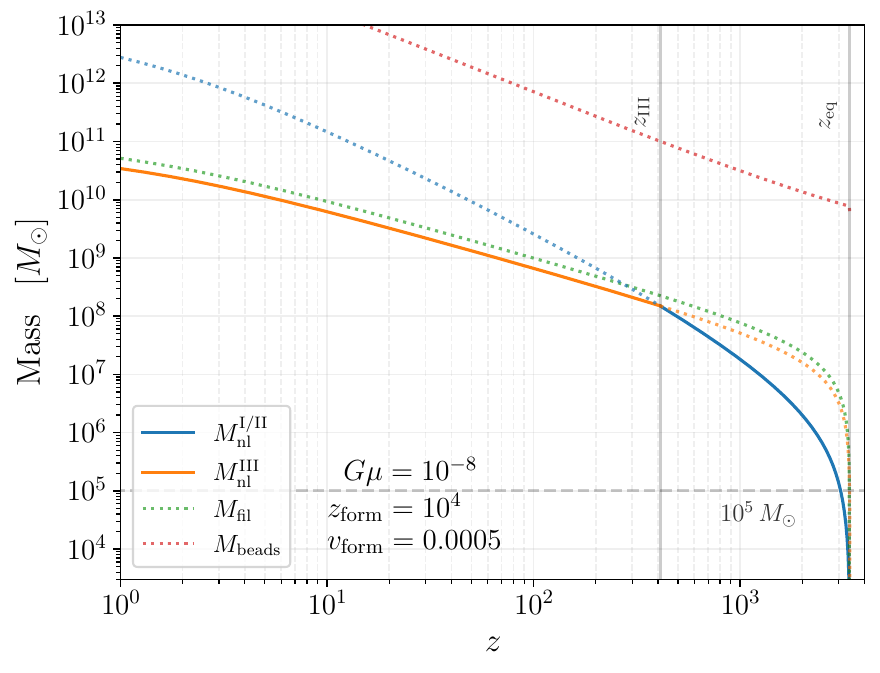}
\hspace{4mm}
\includegraphics[width=\columnwidth]{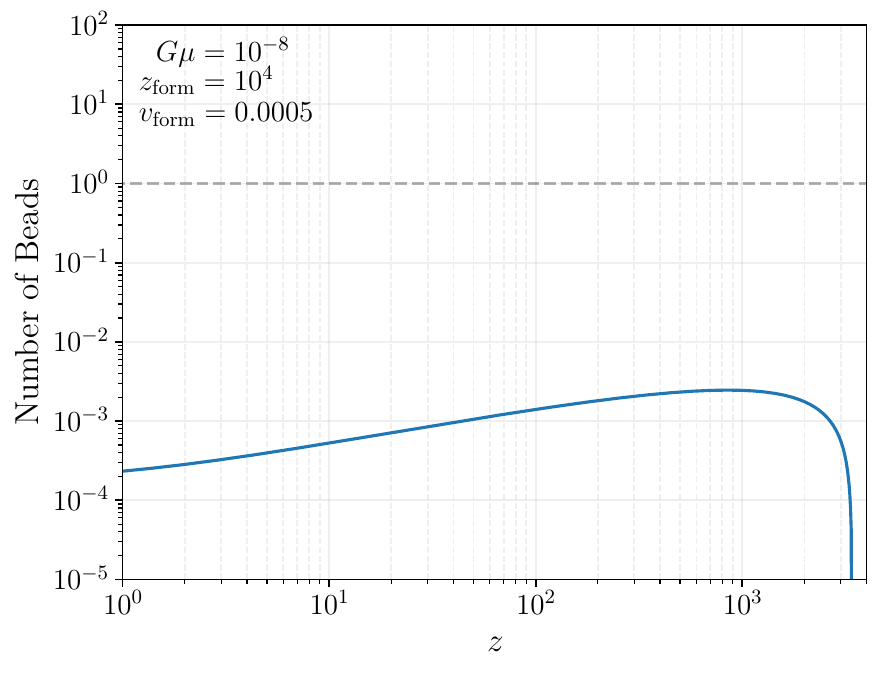}\\
\includegraphics[width=\columnwidth]{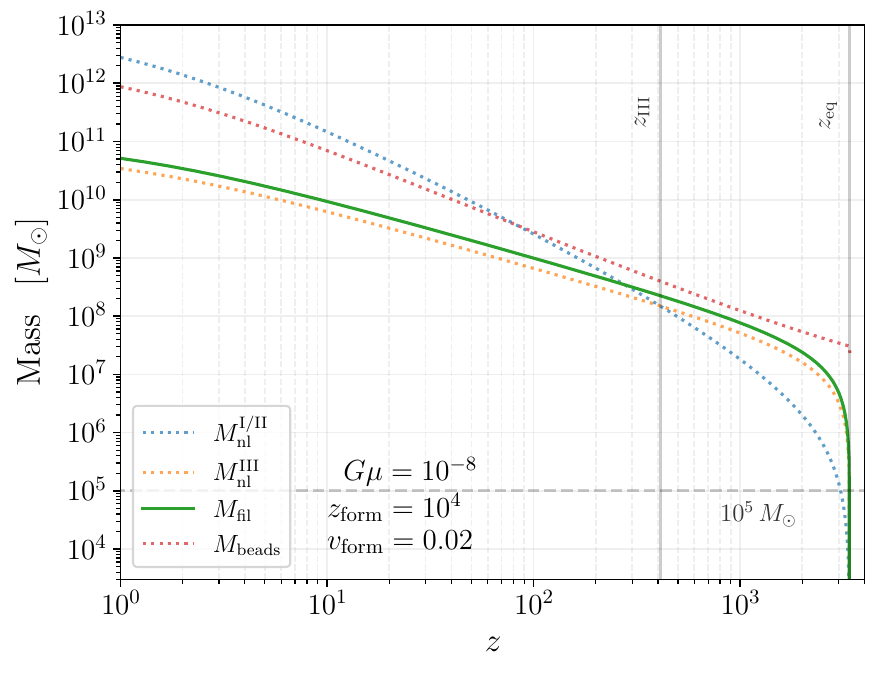}
\hspace{4mm}
\includegraphics[width=\columnwidth]{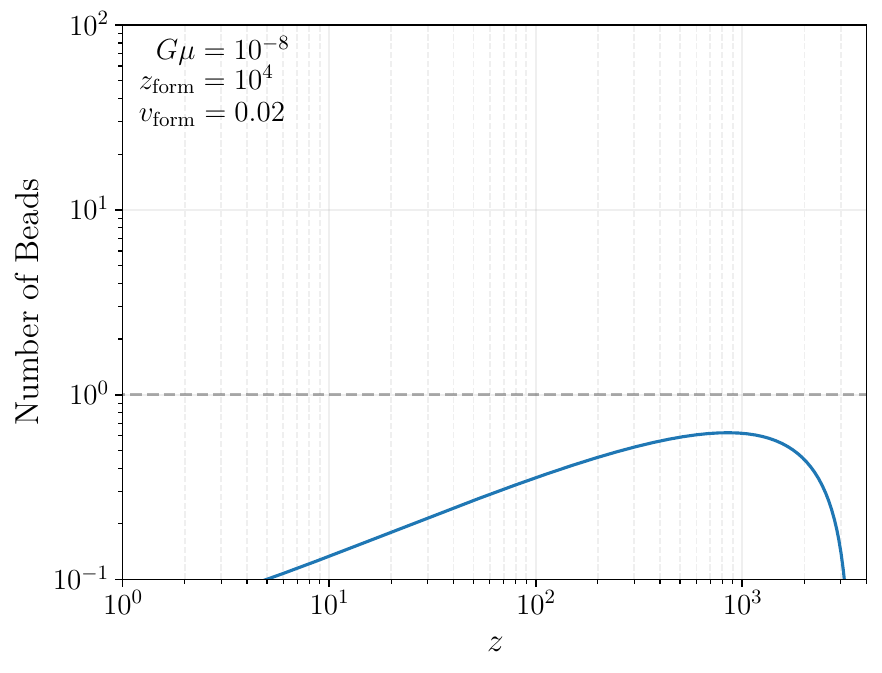}\\
\includegraphics[width=\columnwidth]{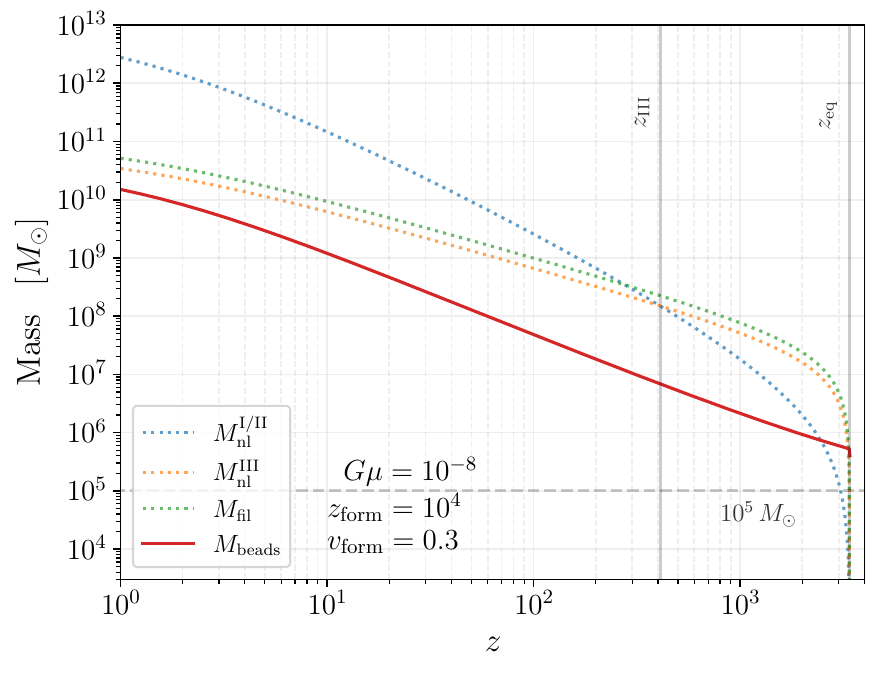}
\hspace{4mm}
\includegraphics[width=\columnwidth]{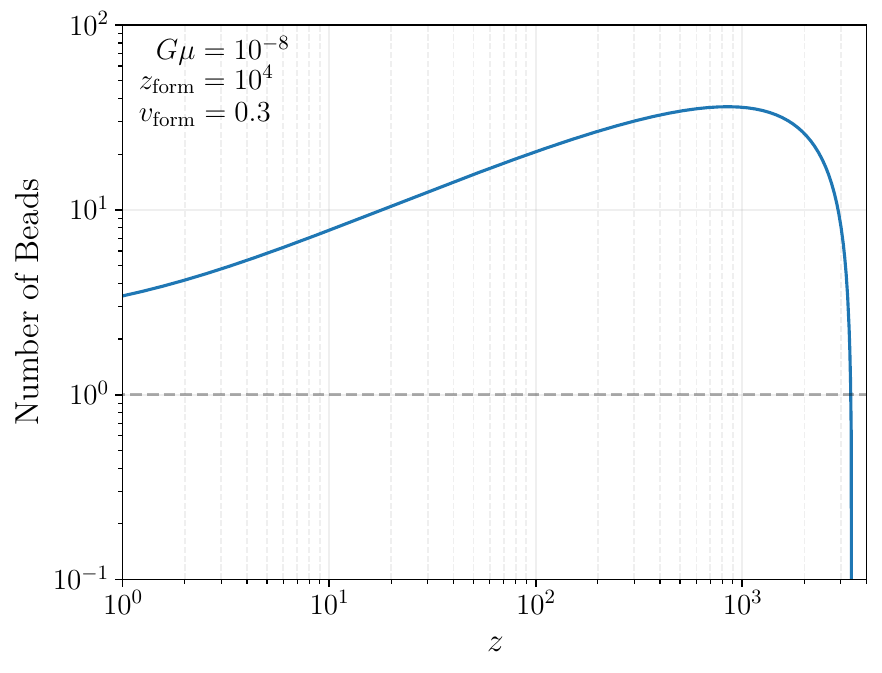}
\caption{From top to bottom, we show the total accreted mass in dark matter halos around string loops with increasingly large velocities relative to the dark matter. We choose benchmark parameters $G\mu = 10^{-8}$, $z_{\rm form} = 10^4$ as they highlight all four potential avenues for growth. In the top panels, the loops move slow enough for the spherical accretion rates to apply, with both Region I/II and Region III accretion playing an important role. In the middle panel, a single long filament grows but does not fragment. In the bottom panel, fragmentation occurs and $\mathcal{O}(10)$ beads are observed to form. In each panel the active accretion paradigm is labeled with the solid line, while the unrealized ones are shown as dotted lines, for comparison.}
\label{Fig:accretion_rates}
\end{figure*}
%
The number of beads at any given redshift need not be a whole number within this approximate scheme, so one should interpret the total number of halos formed from a given string loop at redshift $z$ as $n_{\rm halo,\,loop} = {\rm Max}[1,{\rm Floor}(n_{\rm bead})]$. Perhaps most importantly, the mass of a bead grows with time following $M_{\rm bead, N\geq1} = M_{\rm fil}/n_{\rm bead}$, provided that $n_{\rm bead} \geq 1$. For this we find
%
\begin{align} \label{eq:M_bead}
     M_{\rm bead, N\geq1}  = &4\pi \frac{t_{\rm eq}}{G} \frac{\tilde{M}(z)}{\tilde{L}(z)} \nonumber \\
     &\times \left[ \frac{\alpha G\mu}{v_{\rm form}} \left(\frac{1+z_{\rm eq}}{1+z_{\rm form}} \right) \left(\frac{1+z_{\rm eq}}{1+z} \right)\right]^{3/2}. \nonumber
\end{align}
%
Thus, the true accretion rate of a string-seeded halo moving with a non-negligible velocity with respect to the dark matter is given by
%
\begin{align}
    M_{\rm nl, v}(z) = {\rm Min}(M_{\rm fil},M_{\rm bead, N\geq1} ).
\end{align}
%
Each string loop in this case is capable of sourcing a number of sizeable overdensities given by $n_{\rm halo, loop}$, which is defined above. 

In Fig.~\ref{Fig:accretion_rates} we provide a comparison between the total accreted mass in each different regime as a function of redshift. To highlight the relevance of each effect, we consider a loop forming at $z_{\rm form} = 10^4$ with string tension of $G\mu = 10^{-8}$, and vary its formation velocity. The left hand panels in this figure show the halo mass for a particular loop velocity, while the right-hand panels shows the number of beads expected. In the left panels, the type of accretion that is applicable for a given velocity is labeled using a solid curve, while the unrealized mechanisms are given by the faded dotted lines.

In the top two panels, the slowest possible loops\footnote{Of course, the non-linear mass in this panel accurately describes any loops produced with $v \leq 5 \times 10^{-4}$.} are considered, with $v_{\rm form} = 5 \times 10^{-4}$. In this case, one finds that $z_{\rm sc} > z_{\rm eq}$, thus the spherical collapse formalism can be used to determine the growth rate. For these loops, $M_{\rm nl}^{\rm I/II}$ growth is observed at high redshifts, with a transition to the point mass limit ($M_{\rm nl}^{\rm III}$) around $z_{\rm III} \simeq 400$. Loops that form earlier will have $z_{\rm III}$ closer to matter radiation equality, but the overall halo mass will be suppressed due to the fact that the loop is smaller and thus less massive. Clearly, the number of beads in this scenario is $n_{\rm bead} \ll 1$ implying that no fragmentation occurs.

The middle two panels showcase slightly faster loops, with $1+z_{\rm sc} \leq 0$, implying that the stationary spherical accretion case never applies. At the same time, the loop is not moving fast enough for fragmentation to occur, as is evidenced by the fact that we have $n_{\rm bead} \leq 1$. In this case, the total halo mass is simply the $M_{\rm fil}$, and has a somewhat elongated shape. Nevertheless, we still expect that at late times, the filament will collapse if the Jeans mass is exceeded.

Finally, the bottom two panels showcase the evolution of halos in the fragmented scenario, triggered by the transverse instability mentioned above. Efficient fragmentation requires $v_{\rm form} \gtrsim 0.1$, which as we will see shortly is a rather typical value from string network simulations. In this case, the mass of each individual bead is suppressed, with that suppression growing as the velocity increases. Since the total mass in a filament is velocity independent, however, conservation of mass tells us that $n_{\rm bead}$ rises in response. For $v_{\rm form} = 0.3$, it is typical for a loop to source $\mathcal{O}(10)$ dark matter halos.

\subsubsection{Loop velocity distribution}
As we have seen in the previous subsection, as well as in Fig~\ref{Fig:accretion_rates} the type of accretion a string loop undergoes is highly dependent on its velocity relative to the dark matter. When produced, numerical studies indicate that loops tend to have a fairly wide range of initial velocities. To our knowledge, simulations have not yet bestowed us with a full velocity distribution, but rather seem to indicate that the distribution appears to be peaked around $ v_{\rm form} \simeq 0.3$, with wide tails. In order to estimate the effects of this velocity spread, we posit that a reasonable PDF for the distribution is given by
%
\begin{align} \label{eq:f_v}
    f(v_{\rm form}) = B \, v_{\rm form}^2 (1-v_{\rm form}^2)^p.
\end{align}
%
Phenomenologically, $p = 10$ yields $\langle v_{\rm form} \rangle \simeq 0.3$, and leads to a normalization constant of $B \simeq 85$. In what follows we will use this distribution when computing mass functions discussing abundances of direct collapse black holes.

\begin{figure}
\includegraphics[width=0.95\columnwidth]{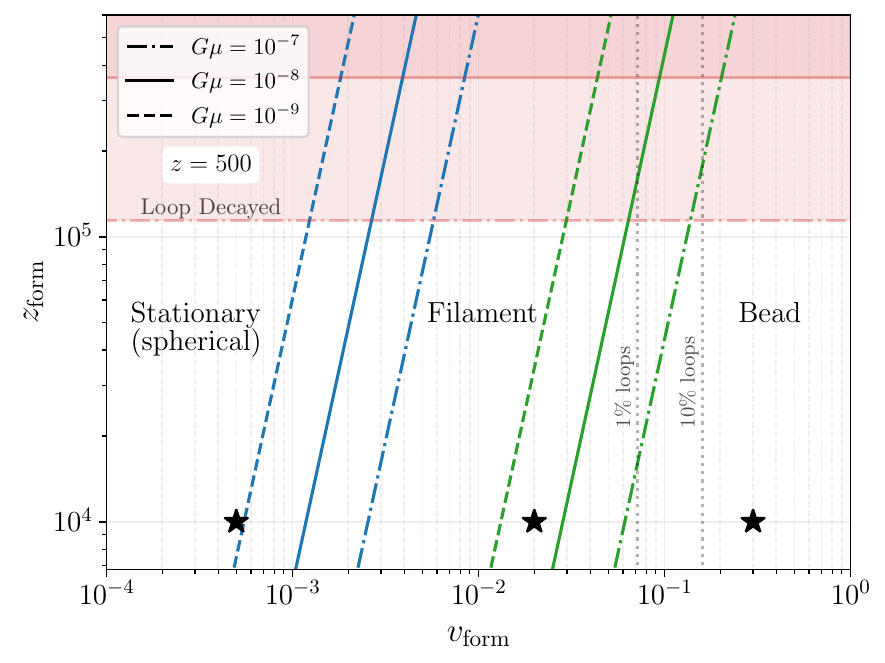}
\caption{Different accretion scenarios for a given $z_{\rm form}$ and $v_{\rm form}$. As the initial loop speed is dialed up, one transitions from stationary (left of the blue curve), to filamentary (between the blue and green curves), and finally fragmentation into beads (right of the green line). For the velocity distribution given in Eq.~\eqref{eq:f_v}, the vast majority of loops produce a number of beads which will then grow independently. Stars indicate parameter choices whose detailed accretion rates are shown in Fig.~\ref{Fig:accretion_rates}.  }
\label{fig:zform_vform_z500}
\end{figure}

As discussed earlier, we are mainly interested in loops forming at $z_{\rm form} \geq 2 z_{\rm eq}$. In Fig.~\ref{fig:zform_vform_z500}, we show the different accretion regimes experienced by a loop for a given $z_{\rm form}$ as a function of the formation velocity. We choose to evaluate the respective accretion conditions at a redshift of $z = 500$, which as we will see in the next section is roughly the time in which the direct collapse of baryons may occur in a string-seeded overdensity.  

In this figure, regions to the left of the blue line experience spherical growth, following Eq.~\eqref{eq:Mnl_all}. Regions between the blue and green contours develop elongated (filamentary) turnaround surfaces without fragmentation, while to the right of the green contour fragmentation into beads will occur. We show how these contours shift for different values of $G\mu = (10^{-7},10^{-8},10^{-9})$, with the dash-dotted, solid, and dashed lines respectively. From left to right, the stars represent benchmark cases that were presented in Fig.~\ref{Fig:accretion_rates} (from top to bottom). At $z=500$, the red horizontal lines indicate at which formation redshift (for a given $G\mu$), the string loop will have fully decayed. We remind the reader that the decay of a string loop does not mean the halo or filament ceases to grow, but instead that these cases need to be handled more carefully. Finally, the two dotted vertical lines highlight the velocities below which only $10\%$ and $1\%$ of loops are produced, following Eq.~\eqref{eq:f_v}. As can be surmised by the 1\% and 10\% contours, the vast majority of loop-seeded overdensities following this distribution exhibit cylindrical fragmentation. 

%
%

\section{Direct collapse of string-seeded overdensities} \label{sec:direct_collapse}
It has been shown that under the right set of criteria, it is possible for baryons in a halo to undergo direct collapse \cite{Haehnelt1993,Umemura1993, Loeb1994, Eisenstein1994,Haiman1999, Oh2001, Bromm2002, Begelman2006, Inayoshi2019}. In these scenarios, collapse is preceded by a period of atomic cooling, and occurs in a monolithic (i.e. unfragmented) fashion. This collapse forms and feeds a central object, shown in various simulations to either be a black hole, or an unstable protostar which typically collapses into a black hole on relatively short timescales. These numerical studies have furnished us with a set of conditions on the halo under which direct collapse is likely to take place:
\begin{itemize}
    \item \textit{Sufficiently hot}: Halos are required to achieve a virial temperature of $T_{\rm vir} \gtrsim 10^4 \, {\rm K}$. Above this temperature, atomic cooling is triggered and signals the onset of the collapsing phase. 
    \item \textit{Sufficiently pristine}: Halos must not be polluted by a significant fraction of heavy elements or molecules. The presence of these pollutants provide additional pathways for the gas to cool, preventing it from reaching the atomic cooling threshold.
    \item \textit{Sufficiently massive}: If a central object is to form, it needs to be accompanied by sufficient material that it is able to accrete. Simulations indicate that the object needs to sustain a rate of $\dot{M} \simeq 0.1 - 1 \, M_{\odot}/{\rm yr}$ for roughly $10^5$ years in order to grow to a so-called heavy seed.
\end{itemize}
Practically speaking, the \textit{pristine} condition actually demands two things. First, the production of heavy elements requires some level of star formation. A trivial (but useful) solution to this condition is to only consider halos at times earlier than the typical redshift of star formation, say $z \gtrsim 20$. Second, to inhibit the creation of molecular hydrogen (${\rm H}_2$), one typically requires a significant flux of Lyman-Werner photons ($10 \, {\rm eV} \lesssim E_{\rm \gamma} \lesssim 13 \, {\rm eV}$) to provide direct dissociation or to disrupt its other formation pathways\footnote{It is precisely this lack of Lyman-Werner background that makes the direct collapse of halos formed from $\Lambda$CDM seeds at $z\simeq 20$ improbable.}. Notably, Qin \textit{et al.} \cite{Qin2025} have recently highlighted the fact that the cosmic microwave background is sufficiently energetic to suppress ${\rm H}_2$ formation at $z \gtrsim 200$ \cite{Tegmark1996, Hirata2006, Qin2023}. 

With this in mind, it becomes clear that dark matter halos present at high redshifts need only to satisfy the \textit{sufficiently hot} and \textit{massive} conditions in order to provide ideal breeding grounds for direct collapse black hole formation. The main aim of this section is to determine precisely which subset of string-seeded overdensities can meet these criteria, and thus produce heavy seeds at early times. 

A final salient fact to keep in mind is the relative streaming velocities between the baryons and dark matter after recombination \cite{Tseliakhovich2010}. One can argue\footnote{See the end matter of \cite{Qin2025} for a back-of-the-envelope estimate of this.} that baryons are unable to fall into potential wells sourced by early overdensities unless the relative velocity between the baryons and dark matter is less than the associated virial velocity of the halo. For atomic cooling halos ($T_{\rm vir} \simeq 10^4 \, {\rm K}$) this occurs at $z \simeq 500$. In what follows (and also in Fig.~\ref{Fig:accretion_rates}), we evaluate our direct collapse conditions and mass functions at this redshift for maximum clarity. Of course, all expressions derived in the main text retain their full time dependence, allowing one to easily translate our results to any redshift.

\subsection{Sufficient temperature}
The virial temperature around a spherical overdensity is well known \cite{virial} to be given by 
%
\begin{align} \label{eq:Tvir}
    T_{\rm vir} = \frac{\mu m_{\rm p} G M_{\rm nl }}{r_{\rm vir}},
\end{align}
%
where $\mu \simeq 0.6$ is the mean molecular weight of hydrogen, and $m_{\rm p}$ is the proton mass. Around a stationary cosmic string loop, one can approximate this by setting $M_{\rm nl} = M_{\rm nl}^{\rm III}$ and $r_{\rm vir} \simeq  r_{\rm nl}/4 = q_{\rm nl}/[4(1+z)]$ to find \cite{Cyr2022}
%
\begin{align} \label{eq:Tvir_pm}
    T_{\rm vir}^{\rm III} \simeq 10^{7} \, {\rm K} &\left(\frac{G\mu}{10^{-8}} \right)^{2/3} \left(\frac{1+z}{1+z_{\rm form}}\right)^{1/3} \nonumber \\
    &\hspace{16mm}\times \left( \frac{1+z_{\rm eq}}{1+z_{\rm form}}\right) \tilde{M}^{2/3}(z).
\end{align}
%
Requiring $T_{\rm vir}^{\rm III} \gtrsim 10^4 \, {\rm K}$ at roughly $z \simeq 500$ provides a rather lax requirement on the string tension and formation redshift. When the loops form with a non-negligible velocity, however, it was shown \cite{Eisenstein1996, Shlaer2012} that the virial temperature can be estimated by replacing $M_{\rm nl} = M_{\rm fil}$ and $r_{\rm vir} = L_{\rm fil}$ in Eq.~\eqref{eq:Tvir}. For $z \ll z_{\rm eq}$, this yields
%
\begin{align} \label{eq:Tvir_fil}
    T_{\rm vir}^{\rm fil} \simeq 10^4 \, {\rm K} &\left( \frac{G\mu}{10^{-8}} \right) \left(\frac{v_{\rm form}}{0.3} \right)^{-1} \nonumber \\
    &\times \left(  \frac{1+z_{\rm eq}}{1+z_{\rm form}}\right) \frac{\tilde{M}(z)}{\tilde{L}(z)}.
\end{align}
%
Clearly, this leaves less room to adjust the model parameters while still satisfying $T_{\rm vir} \geq 10^4 \, {\rm K}$, though loops in the low velocity tail of the distribution will still tend to produce sufficiently hot halos. Aside from $\tilde{M}/\tilde{L}$ (which yields a small enhancement at early times), the virial temperature of a filament is redshift independent. In deriving Eq.~\eqref{eq:Tvir_fil} Eisenstein, Loeb and Turner \cite{Eisenstein1996} assumed an infinitely straight, isothermal cylinder with constant mass density. Note that this simplified picture neglects the fragmentation of the cylinder, and one may expect that the virial temperature within any individual bead could be higher than that given by $T_{\rm vir}^{\rm fil}$ as the beads form and subsequently grow in a quasi-spherical way. We leave a more comprehensive analysis of the virial temperature within a bead to future work, noting that Eq.~\eqref{eq:Tvir_fil} provides a conservative lower bound when assuming $T_{\rm vir}^{\rm bead} = T_{\rm vir}^{\rm fil}$. Under this assumption we have the requirement
%
\begin{align} \label{eq:virial_v}
    v^{\rm vir}_{\rm form} \leq 0.3 \left( \frac{G\mu}{10^{-8}}\right) \left(\frac{1+z_{\rm eq}}{1+z_{\rm form}} \right) \frac{\tilde{M}(z)}{\tilde{L}(z)},
\end{align}
%
in order for the halo to attain the required temperature to undergo atomic cooling. This condition holds for both the beaded and filamentary scenarios. Extraordinary low velocity loops ($v_{\rm form} \lesssim 5 \times 10^{-4}$) would need to be analyzed by applying the virial condition on Eq.~\ref{eq:Tvir_pm}. With the velocity distribution given in Eq.~\eqref{eq:f_v}, the fraction of loops satisfying this is $\simeq 10^{-9}$, thus we can safely neglect this population of slow movers.

\subsection{Sufficient mass}
Considering now only loops with an appreciable velocity, the second condition on the string-seeded halo is that sufficient mass ($M_{\rm nl} \gtrsim 10^5 \, M_{\odot}$) is present at the time of direct collapse. Satisfying this condition depends on whether the overdensity is filamentary, or fragmented, so we present both cases. In the case of fragmentation into beads, the growth of a fragmented halo is given by Eq.~\eqref{eq:M_bead}, which gives the following condition on the velocity
%
\begin{align}
    v_{\rm form}^{\rm mass} \leq 1.44 \left( \frac{G\mu}{10^{-8}}\right) &\left(\frac{1+z_{\rm eq}}{1+z_{\rm form}} \right) \left(\frac{\tilde{M}(z)}{\tilde{L}(z)}\right)^{2/3} \nonumber \\
    &\hspace{17mm}\times \left( \frac{1+z_{\rm eq}}{1+z}\right).
\end{align}
%
An examination of this condition reveals that it is in fact degenerate with Eq.~\eqref{eq:virial_v}. The larger prefactor and the $(1+z)^{-1}$ dependence ensure that if the virial condition is met, each bead will have more than enough mass to grow the central black hole/protostar.

When fragmentation does not occur, the filament remains connected but grows to be highly elongated. As we saw earlier, the total mass of the overdense cylinder is velocity independent, thus one can instead rephrase the \textit{sufficient mass} condition in this case to one on the formation redshift. By requiring that $M_{\rm fil} \geq 10^5 \, M_{\odot}$, we find
%
\begin{align}
    \left(\frac{1+z_{\rm form}}{1+z_{\rm eq}} \right) \leq 50\, \left[\tilde{M}(z) \left( \frac{G\mu}{10^{-8}}\right) \left(\frac{1+z_{\rm eq}}{1+z}\right) \right]^{1/2}.
\end{align}
%
Typically speaking, halos and filaments generated by loops forming at $z_{\rm form} \gtrsim 10^2 z_{\rm eq}$ will be unable to host a direct collapse black hole. Usually, however, the sufficient mass constraint is superseded by the virial constraint discussed in the previous subsection.

\subsection{Abundance of direct collapse black holes}
We are now in a position to consistently determine the abundance of string seeded halos. The full mass function of halos is given by
%
\begin{align} \label{eq:halo_mass_function}
    \frac{\id N_{\rm halo}}{\id M} = \int_0^1\id v_{\rm form} \, f(v_{\rm form}) \, n_{\rm halo,\,loop}  \frac{\id N}{\id L} \frac{\partial L}{\partial M},
\end{align}
%
where $f(v)$ is the normalized velocity distribution function, $\id N/\id L$ is the spectrum of loops, and $\partial L/\partial M$ is the Jacobian transformation taking us to whichever accretion scenario is relevant for a given choice of model parameters. The Jacobian factors are given explicitly in Appendix~\ref{sec:Jacobian_Factor}. As described above, $n_{\rm halo,loop} \geq 1$ counts the number of halos sourced by a single loop. 

\begin{figure*}
\centering 
\includegraphics[width=\columnwidth]{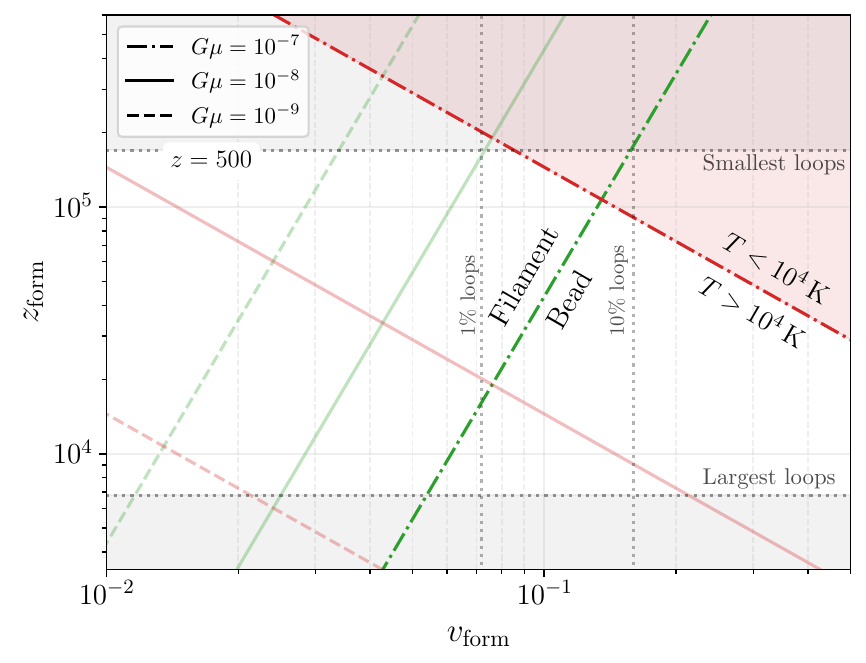}
\hspace{4mm}
\includegraphics[width=\columnwidth]{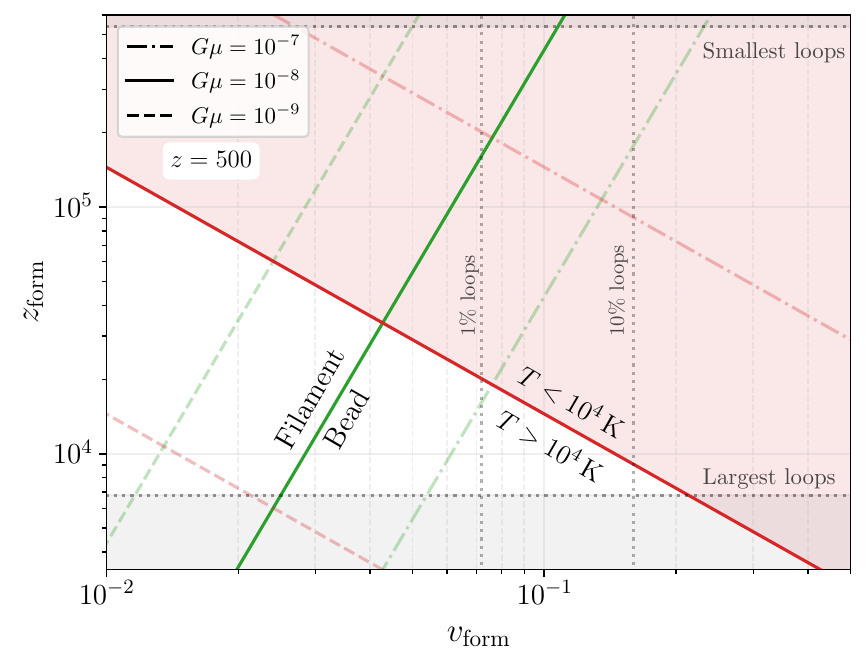}\\
\includegraphics[width=\columnwidth]{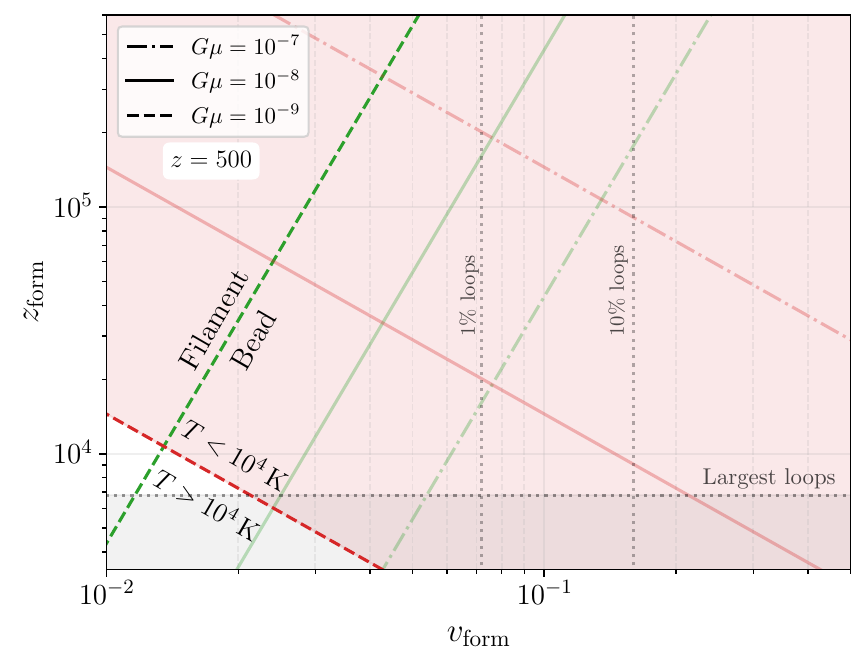}
\hspace{4mm}
\includegraphics[width=\columnwidth]{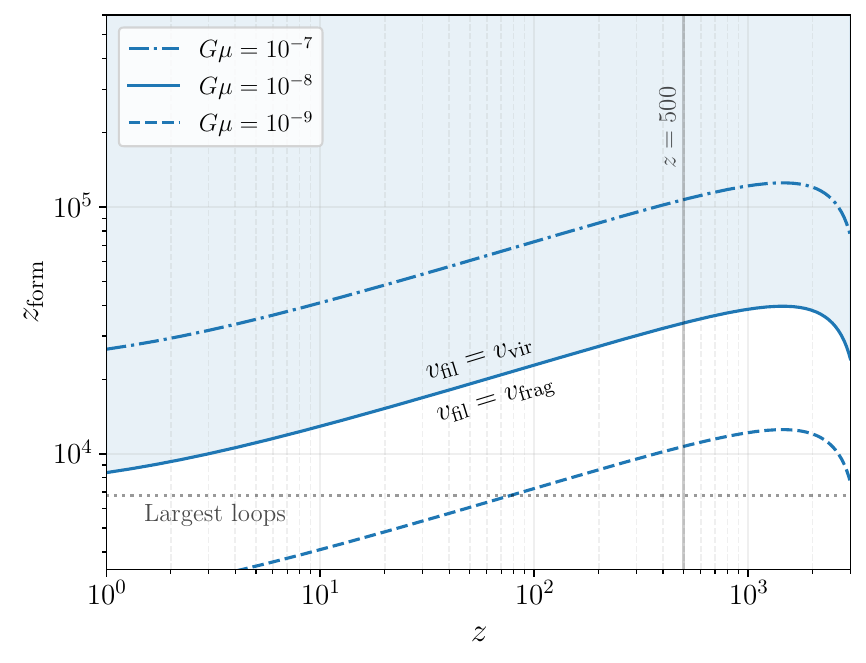}
\caption{The relationship between loop formation redshift and velocity. The prominent contours in the top left, top right, and bottom left panels correspond to $G\mu = (10^{-7},10^{-8},10^{-9})$ respectively. The red shaded region indicates the part of parameter space in which a string-seeded overdensity does not meet the atomic cooling threshold. To the right of the green contour ($v \geq v_{\rm form}^{\rm frag}$) fragmentation of cylindrical overdensities will occur. The intersection between the fragmentation and velocity curves always occurs at $M = M_*$. Bottom right: above the blue curve fragmented cylinders will never produce beads with $T > 10^4 \, {\rm K}$, whereas below one will find hot filaments and beads.}
\label{Fig:conditional_vir_contours}
\end{figure*}

A subset of these halos will also be both heavy enough and hot enough to host direct collapse black holes. We are most interested in their mass function, given by
%
\begin{align} \label{eq:dndM_total}
    \frac{\id N_{\rm Halo,BH}}{\id M} = \frac{\id N}{\id M_{\rm pm}} + \frac{\id N}{\id M_{\rm fil}} + \frac{\id N}{\id M_{\rm bead}}.
\end{align}
%
For typical velocity distributions (like the one considered in Eq.~\eqref{eq:f_v}), the fraction of loops moving slow enough to accrete in the stationary point mass limit is negligible, thus we neglect the first term. Consider now the second term, which comes from filaments that do not fragment. A reasonable approximation to the mass function is given by
%
\begin{align} \label{eq:dNdMfil}
    \left.\frac{\id N}{\id M_{\rm fil}}\right|_{z_{\rm eq}} &= \int_0^{v_{\rm fil}} \id v_{\rm form} \, f(v_{\rm form}) \, n_{\rm halo, loop} \frac{\id N}{\id L} \frac{\partial L}{\partial M_{\rm fil}} \nonumber \\
    &= \frac{\id N}{\id L} \frac{\partial L}{\partial M_{\rm fil}} \int_0^{v_{\rm fil}} \id v_{\rm form} f(v_{\rm form}).
\end{align}
%
Where $v_{\rm fil} = {\rm Min}(v_{\rm form}^{\rm frag}, \,v_{\rm form}^{\rm vir})$ is determined by either the velocity below which the virial condition in Eq.~\eqref{eq:virial_v} is met, or the velocity above which fragmentation occurs. One can derive the fragmentation velocity by setting $n_{\rm bead} = 1$ in Eq.~\eqref{eq:N_bead}, which yields
%
\begin{align}
    v_{\rm form}^{\rm frag} \simeq  \frac{8\times 10^{-3}}{\tilde{L}^{2/3}(z)}\left(\frac{G\mu}{10^{-8}} \right)^{1/3} \left(\frac{1+z_{\rm form}}{1+z} \right)^{1/3}.
\end{align}
%
In Fig.~\ref{Fig:conditional_vir_contours}, we compare this fragmentation velocity (green curve) with the virial velocity (red curve) given in Eq.~\eqref{eq:virial_v}. Recall that $z_{\rm form}$, the loop size ($L$), and the halo mass can all be directly related to each other using expressions found in Appendix~\ref{sec:Jacobian_Factor}. For cylindrical overdensities, there is an unambiguous relationship between the formation redshift and the filament mass. The most prominent contours correspond to string tensions of $G\mu = (10^{-7}, 10^{-8},10^{-9})$ in the top left, top right, and bottom left panels respectively.

%
%

Regions shaded in red indicate which parts of the parameter space the virial temperature of an overdensity is $T < 10^4 \, {\rm K}$, thus atomic cooling cannot occur. We note that this does not preclude the formation of dark matter halos, merely that halos forming from smaller and faster loops will tend to be rather cool. The green contour demarcates the formation velocities in which fragmentation is expected to occur, with beads forming to the right of this line. 

The gray dotted lines mark the maximum ($z_{\rm form} = 2z_{\rm eq}$) and minimum ($z_{\rm decay} \simeq z_{\rm eq}/2$) loop sizes considered in this work, with further details found in Appendix~\ref{sec:Mass_range}. These gray shaded regions should not be thought of as hard cutoffs to the corresponding mass function, but rather regions of parameter space in which a more refined analysis is required, which we defer to future work.

The bottom right panel shows contours of $v_{\rm form}^{\rm vir} = v_{\rm form}^{\rm frag}$ for a given string tension, as a function of redshift. In the blue shaded regions, beads produced by fragmented filaments will never be hot enough to host a direct collapse black hole. Note that the intersection of the vertical ``$z = 500$" line with any of these blue contours corresponds exactly with the intersection of the green and red curves in the other panels. The mass scale that corresponds to this intersection is given by
%
\begin{align}
    M_* \simeq 10^6 \, M_{\odot} \, \left( \frac{1+z_{\rm eq}}{1+z}\right)^{3/2} \left(\frac{\tilde{L}(z)}{\tilde{M}(z)} \right)^{1/2}.
\end{align}
%
Interestingly, this critical mass is dependent only on redshift. 

Fig.~\ref{Fig:conditional_vir_contours} is highly informative, as it allow us to visualize the integration bound in a straightforward manner. Moving from left to right along the $v_{\rm form}$ axis, the upper bound on the filament mass function will be given by either the virial or the fragmentation velocity, whichever is encountered first. The least massive string loops (those forming at high redshift) will only produce filaments that are hot enough to host a DCBH. More massive halos will instead possess both hot filaments and beads. The transition between these two regimes happens at exactly $M_{\rm fil} = M_*$.
%
\begin{figure*}
\centering 
\includegraphics[width=\columnwidth]{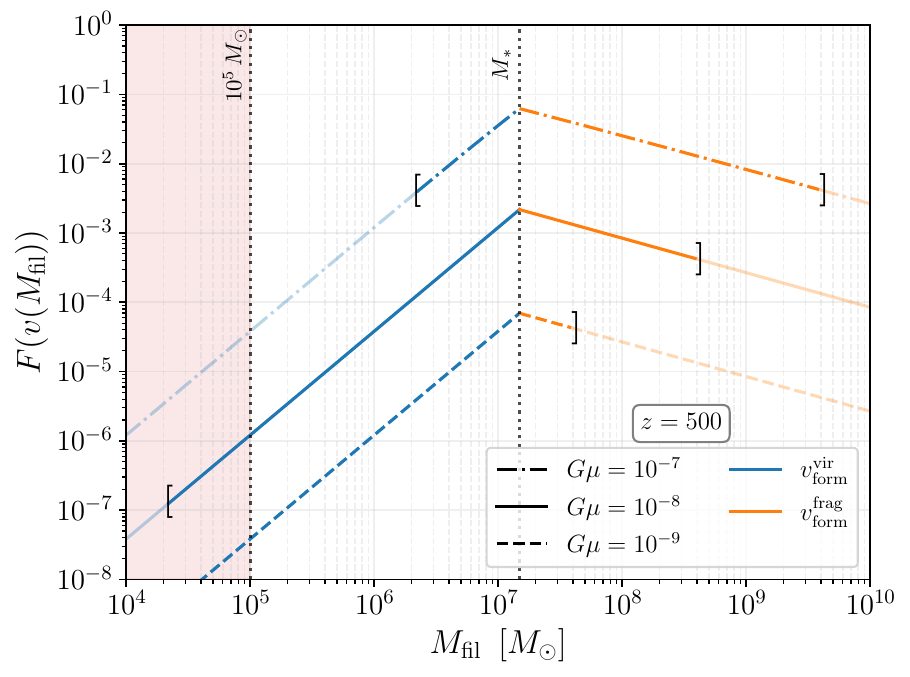}
\hspace{4mm}
\includegraphics[width=\columnwidth]{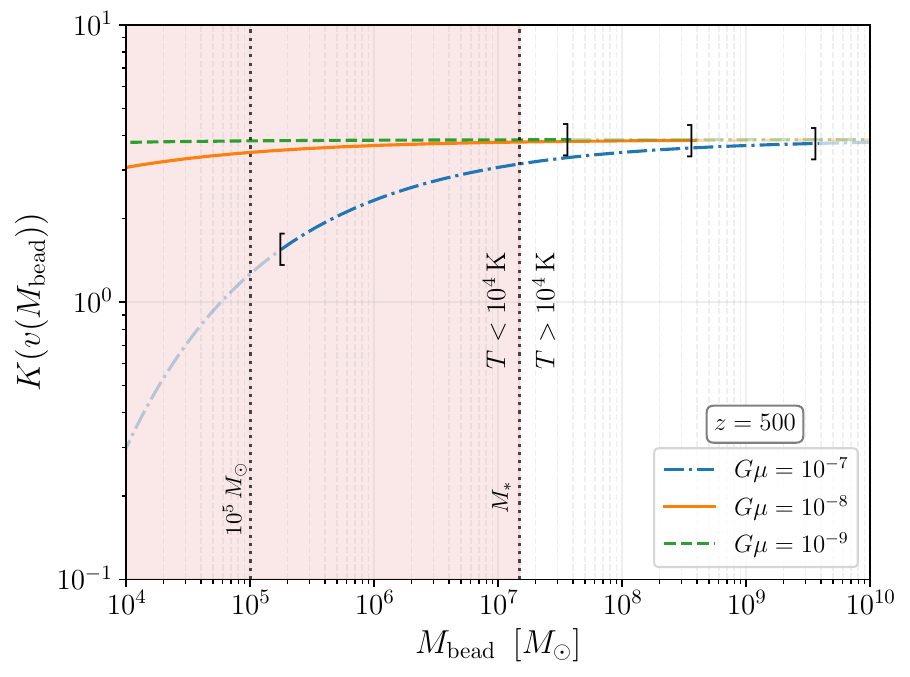}
\caption{Velocity distribution function integrals given in Eqs.~\ref{Eq:F_v_int} (left panel) and \ref{Eq:K_v_int} (right panel). Brackets represent the minimum and maximum halo masses for a particular $G\mu$ value. Red shaded regions represent the parameter space in which at least one of the direct collapse criteria is not met. }
\label{Fig:distribution_probs}
\end{figure*}
%

Let us now return to the determination of the filament mass function, Eq.~\eqref{eq:dNdMfil}. Neither the loop number density nor the Jacobian factor depend on velocity, so we are free to pull them outside of the integral. In addition to this, $n_{\rm halo,loop} = 1$ in the filament scenario. At this point, the integration over $f(v)$ may now be performed with no additional subtleties. For completeness, we use the number density of loops evaluated at matter radiation equality, as this provides the initial conditions for the spectrum of overdensities. Explicitly, we use
%
\begin{align}
    \frac{\id N}{\id L} &= \frac{\alpha_{\rm r}}{t_{\rm eq}^{3/2}(L + \Gamma_{\rm g} G\mu t_{\rm eq})^{5/2}} \nonumber \\
    &\approx \frac{\alpha_{\rm r}}{t_{\rm eq}^{3/2}} \left(\frac{3 G\mu}{5 G} \left[\frac{1+z_{\rm eq}}{1+z} \right] \frac{\tilde{M}(z)}{M_{\rm fil}}\right)^{5/2}.
\end{align}
%
In going to the second line, we have written $L = L(M_{\rm fil})$ which can be readily derived by integration of the Jacobian factor given in Appendix \ref{sec:Jacobian_Factor}. We have also neglected the term proportional to the $\Gamma_{\rm g}$, as we will only be considering loops which survive for at least a Hubble time after recombination. For those loops, $L \gg \Gamma_{\rm g} G\mu t_{\rm eq}$ and the approximation is justified. In this setup, the Jacobian factor then acts as a transfer function which evolves the total halo mass to later redshifts. The central object we wish to compute is the comoving mass function, which is given by
%
\begin{align}
    \left.\frac{\id N}{\id M_{\rm fil}}\right|_{\rm c} = \frac{1}{(1+z_{\rm eq})^{3}} \left.\frac{\id N}{\id M_{\rm fil}}\right|_{z_{\rm eq}}.
\end{align}
%
If we now define
%
\begin{align} \label{Eq:F_v_int}
    F(v_{\rm fil}) \equiv \int_0^{v_{\rm fil}} \id v_{\rm form} \, f(v_{\rm form}),
\end{align}
%
we can then perform some simple algebra to find the relatively tidy expression
%
\begin{align} 
    \left.\frac{\id N}{\id M_{\rm fil}}\right|_{\rm c} &= \frac{\alpha_{\rm r}}{t_{\rm eq}^3}\left[\frac{3t_{\rm eq}}{5G} \frac{G\mu}{\tilde{M}(z)}\left(\frac{1+z_{\rm eq}}{1+z} \right) \right]^{3/2} \frac{M_{\rm fil}^{-5/2}}{(1+z_{\rm eq})^3} \nonumber \\
    &\times \bigg[F(v_{\rm form}^{\rm vir})\Theta(M_*-M_{\rm fil})\Theta(M_{\rm fil} - 10^5 \, M_{\odot}) \nonumber\\
    &\,\,\,+ F(v_{\rm form}^{\rm frag})\Theta(M_{\rm fil}-M_*) \bigg]. 
\end{align}
%
The Heaviside functions in the second line enforce the DCBH conditions for loops forming at high redshifts, while the third line provides the case for late forming loops with $M_{\rm fil} > M_*$. Inserting values yields the perhaps more enlightening form
%
\begin{align}\label{eq:dNdMfil}
    \left.\frac{\id N}{\id M_{\rm fil}}\right|_{\rm c} &= \frac{1.4 \times 10^{14}}{(1+z)^{3/2}\tilde{M}(z)^{3/2}}  \left( \frac{G\mu}{10^{-8}}\right)^{3/2} \left( \frac{M_{\rm fil}}{M_{\odot}}\right)^{-5/2} \nonumber \\
    &\times \bigg[F(v_{\rm form}^{\rm vir})\Theta(M_*-M_{\rm fil})\Theta(M_{\rm fil} - 10^5 \, M_{\odot}) \nonumber\\
    &\,\,\,+ F(v_{\rm form}^{\rm frag})\Theta(M_{\rm fil}-M_*) \bigg] M_{\odot}^{-1} \, {\rm Mpc}^{-3}. 
\end{align}
%
We note that these results are roughly consistent with the mass function inferred by Hao \textit{et al.} \cite{Jiao2024} in the (perhaps unlikely) limit assumed by these authors that all loops are slow moving ($F(v_{\rm form}^{\rm frag}) = 0$ and $F(v_{\rm form}^{\rm vir} )= 1$, and $M_* \rightarrow \infty$). 

We show the scaling of $F(v_{\rm fil})$ as a function of the filament mass in the left hand panel of Fig.~\ref{Fig:distribution_probs}. As discussed, light filaments (stemming from light loops) are bound by the condition on the virial velocity, shown as the blue contours satisfying $M_{\rm fil} \leq M_*$. The orange contours to the right showcase the fragmentation velocity bound. The shape and power-law break of $F(v)$ can be identified with the green and red lines in Fig.~\ref{Fig:conditional_vir_contours}. The brackets indicate the minimum and maximum halo masses for a given $G\mu$ value. Filaments in the red shaded region with $M_{\rm fil} < 10^5 \, M_{\odot}$ do not satisfy all of the DCBH criteria.

A similar analysis can now be performed for the fragmented (or beaded) scenario. We are now required to compute
%
\begin{align} \label{eq:dNdMbead}
    \left.\frac{\id N}{\id M_{\rm bead}}\right|_{\rm c} &= \int_{v_{\rm form}^{\rm frag}}^{1} \id v_{\rm form} \, f(v_{\rm form}) \, n_{\rm halo, loop}  \nonumber \\
    &\hspace{10mm}\times \frac{1}{(1+z_{\rm eq})^3} \frac{\id N}{\id L} \frac{\partial L}{\partial M_{\rm bead}}.
\end{align}
%
The upper bound on this integral goes to $1$ instead of $v_{\rm form}^{\rm vir}$ because there is no longer an unambiguous relationship between $z_{\rm form}$ and $M_{\rm bead}$. In other words, a loop forming at some $z_{\rm form}$ can produce a heavy bead if $v_{\rm form} \gtrsim v_{\rm form}^{\rm frag}$ or a light bead if $v_{\rm form} \simeq 1$. Thus, we are required to integrate over all velocities in which fragmentation occurs. One can show that the subset of loops satisfying $T_{\rm vir} \geq 10^4 \, {\rm K}$ is simply given by $M_{\rm bead} \geq M_*$.  

With this in mind, we can once again perform some algebra to find that the integrand is proportional to $v_{\rm form}^{-1}$. It is then prudent to define a new function
%
\begin{align} \label{Eq:K_v_int}
    K(v_{\rm form}^{\rm frag}) \equiv \int_{v_{\rm form}^{\rm frag}}^{1} \id v_{\rm form} \frac{f(v_{\rm form})}{v_{\rm form}} \nonumber,
\end{align}
%
in order for us to cleanly write the mass function for the beaded regime.
%
\begin{align}
    \left.\frac{\id N}{\id M_{\rm bead}}\right|_{\rm c} &= \frac{1}{4} \frac{\tilde{M}(z)^{5/3}}{\tilde{L}(z)^{2/3}} \frac{(G\mu)^2}{t_{\rm eq}^3} \left(\frac{t_{\rm eq}}{G} \right)^{5/3} \left(\frac{1+z_{\rm eq}}{1+z} \right)^2 \nonumber\\
    &\times\frac{M_{\rm bead}^{-8/3}} {(1+z_{\rm eq})^3} K(v^{\rm frag}_{\rm form} ) \,\Theta(M_{\rm bead} - M_{*}). 
\end{align}
%
Some of the numerical factors have been evaluated to obtain the $1/4$ coefficient. We remind the reader that for the beaded scenario, the \textit{sufficient mass} condition is always satisfied when the virial condition is met. Inserting units once again provides a more insightful form
%
\begin{align} \label{eq:dNdMbead}
    \left.\frac{\id N}{\id M_{\rm bead}}\right|_{\rm c} &= \frac{1.9 \times 10^{14}}{(1+z)^{2}} \frac{\tilde{M}(z)^{5/3}}{\tilde{L}(z)^{2/3}} \left(\frac{G\mu}{10^{-8}} \right)^{2} \left( \frac{M_{\rm bead}}{M_{\odot}}\right)^{-8/3} \nonumber\\
    &\times   K(v_{\rm form}^{\rm frag})\, \Theta(M_{\rm bead} - M_*) \,\,M_{\odot}^{-1} \, {\rm Mpc}^{-3}.
\end{align}
%
We show $K(v)$ as a function of the bead mass in the right hand panel of Fig.~\ref{Fig:distribution_probs}. Unlike $F(v)$ which suppressed the mass function for filaments, this term provides a mild enhancement of the order $K(v) \simeq 4$ for heavy halos. This makes sense, as the velocity distribution we have chosen in Eq.~\ref{eq:f_v} peaks around $v_{\rm form} \simeq 0.3$, where most loops will tend to produce beads instead of filaments. Here, we require $M_{\rm bead} \geq M_* \simeq 10^7 \, M_{\odot}$ in order to satisfy the virial condition.

Eqs.~\eqref{eq:dNdMfil} and \eqref{eq:dNdMbead} are simple analytic expressions which accurately describe the comoving mass function for filaments and beads satisfying the direct collapse black hole conditions. 

One particularly useful aspect of these expressions is that the incorporation of a different $f(v_{\rm form})$ only requires one to recompute the $F(v)$ and $K(v)$ functions. For example, Shlaer \textit{et al.} \cite{Shlaer2012} make the assumption that $f(v) = \delta(v_{\rm form} - 0.3)$. With this, we find $F(v_{\rm fil}) = 0$ and $K(v_{\rm form}^{\rm frag}) = 10/3$, which reproduces the ``normal growth" mass function derived by these authors. Thus, our expressions provide a generalization of the mass functions to any loop velocity distribution in a straightforward way.
%
\begin{figure}
\centering 
\includegraphics[width=\columnwidth]{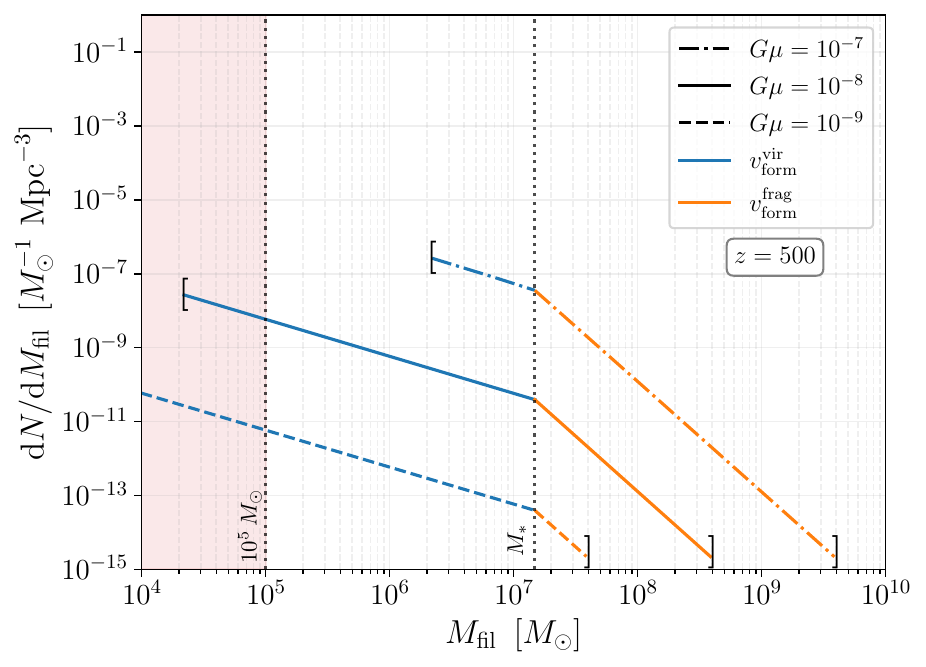}\\
\includegraphics[width=\columnwidth]{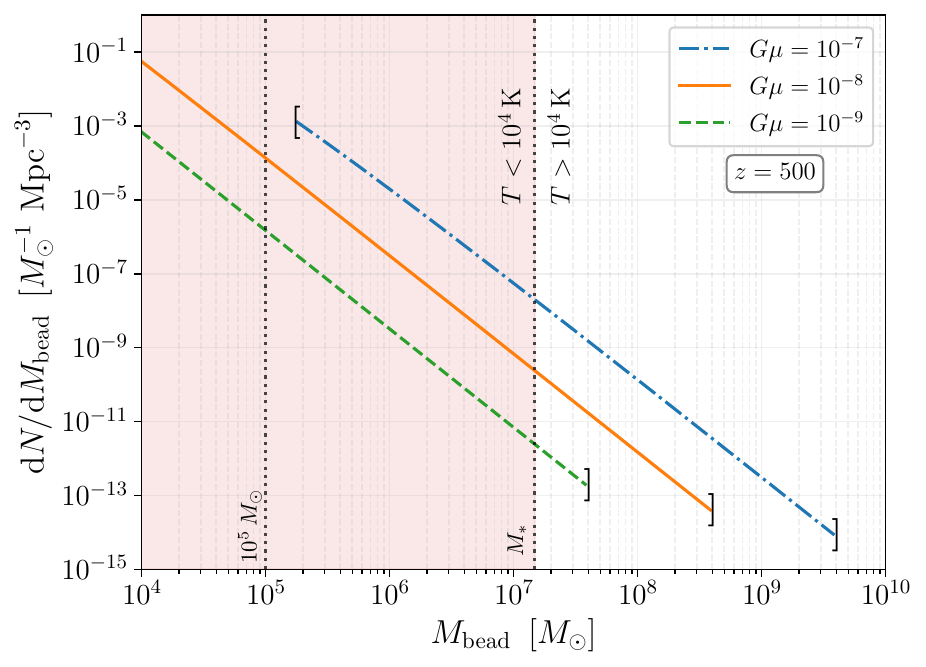}\\
\includegraphics[width=\columnwidth]{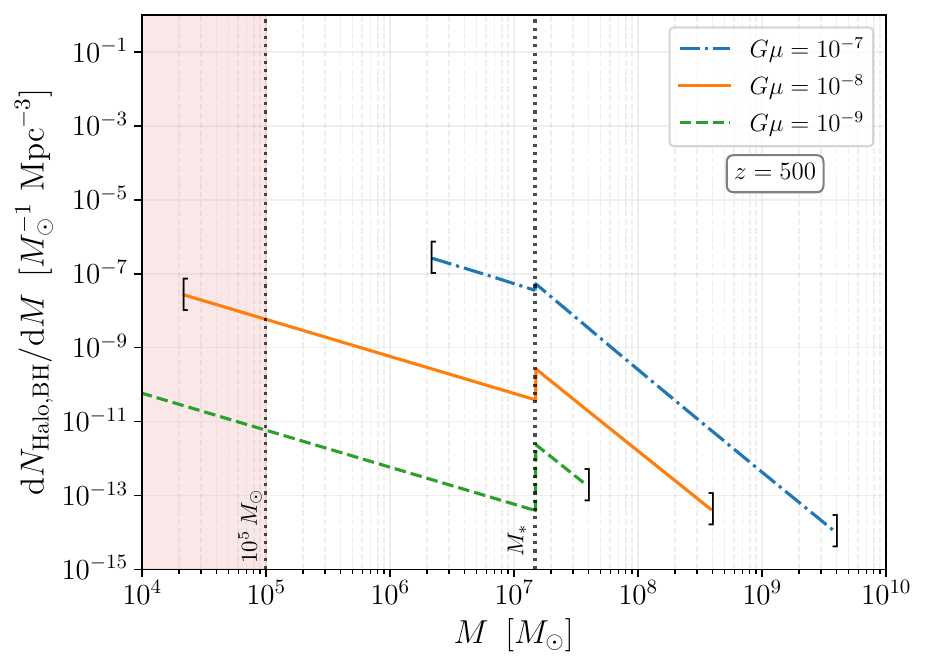}
\caption{From top to bottom, we show the mass functions for filaments [Eq.~\eqref{eq:dNdMfil}], beads [Eq.~\eqref{eq:dNdMbead}], and the total population which could host direct collapse black holes [Eq.~\eqref{eq:dndM_total}].}
\label{Fig:mass_functions}
\end{figure}
%

We present these mass functions in Fig.~\ref{Fig:mass_functions}. The top and middle plots showcase the expected comoving number density of filaments and beads respectively, for different values of $G\mu$. As expected, higher string tensions lead to more massive dark matter halos, albeit over a tighter mass range. This is due to the fact that as one raises $G\mu$, loops decay faster, and thus at matter radiation equality there are simply fewer loops around that formed at high redshift. In fact, observing a somewhat sharp lower cutoff in the mass function at high redshifts would be a smoking gun signature of string-seeded dark matter halos. If detected, the exact position of the cutoff could easily be used to infer the value of $G\mu$.

The power law break at $M_*$ for the filament mass function comes entirely from the piecewise continuous nature of $F(v)$. It's worth noting that it is possible for a filament to have $M<M_*$, while still satisfying $T_{\rm vir} > 10^4 \, {\rm K}$. This is in contrast to the beaded scenario, which  requires $M>M_*$ for any beads wishing to trigger atomic cooling. Though difficult to see, the $G\mu = 10^{-7}$ curve in the bead mass function also inherits some curvature from the corresponding $K(v)$ dependence on $M_{\rm bead}$. 

The bottom panel corresponds to the total mass function of halos capable of hosting direct collapse black holes, schematically given in Eq.~\eqref{eq:dndM_total}. Practically speaking, this mass function is constructed by summing the filament mass function with the sub-population of beaded halos satisfying $M_{\rm bead} \geq M_*$. The hard cutoff imposed at $M_*$ is the reason for the piecewise nature of the total mass function.

Providing the mass functions for string-seeded overdensities satisfying the direct collapse conditions was the primary aim of this work, but we can go a bit further by making some assumptions about the procedure of black hole formation. Heuristically speaking, baryons will begin to fall into the string seeded halos at around $z=500$, after which they will virialize. The virialization procedure takes roughly a free-fall time, given by
%
\begin{align}
    t_{\rm ff} \simeq \sqrt{ \frac{3\pi}{32 G \rho_{\rm bg}}} \simeq \frac{\pi}{2} H^{-1} \nonumber.
\end{align}
%
In terms of redshift, this provides us with a virialized halo of baryons at roughly $z \simeq 225$, at which point the direct collapse of the halo begins. Provided that the halo remains at the atomic cooling threshold throughout the collapse, a central protostar is quickly formed, with simulations indicating it grows at a hypereddington rate of $\dot{M} \simeq 1 \, M_{\odot}/{\rm yr}$ \cite{Inayoshi2014, Inayoshi2019, Umeda2016, Woods2017}. These same simulations indicate that hypereddington growth is terminated once the protostar reaches a mass of $M_{\rm ps} \simeq 10^5 \, M_{\odot}$, after which it collapses into a black hole of similar size. Subsequent growth of the direct collapse black hole occurs through more well-understood channels, which for the purposes of this estimate we will take to be a fraction of the Eddington rate. Therefore, the number density of direct collapse black holes is simply given by the number density of halos satisfying the direct collapse criteria, in other words, an integral over the mass functions presented above. 

In this framework, the initial source population of direct collapse black holes is monochromatic ($M_{\rm form}^{\rm DCBH} \simeq 10^5 \, M_{\odot}$). This is in contrast to the assumption made in \cite{Qin2023}, that the DCBH mass at formation is given by some $\mathcal{O}(1)$ fraction of the halo mass. The variation in high redshift black hole masses inferred by observation thus stem from the specific accretion conditions in each galaxy (e.g. a variation accretion geometries, duty cycles, or other difficult to model effects). We are then interested in calculating
%
\begin{align}
    N_{\rm DCBH} = \epsilon_{\rm L} \int_{{\rm Max}(M_{\rm min}, 10^5 \, M_{\odot})}^{M_{\rm max}} \frac{\id N_{\rm Halo, BH}}{\id M} \,\id M,
\end{align}
%
where $\epsilon_{\rm L}$ is a suppression factor dependent on the angular momentum distribution of string seeded halos. Although not discussed in detail above, direct collapse does require the angular momentum of a halo to be sufficiently low \cite{Bromm2002,Koushiappas2003,Lodato2006}. For typical gaussian perturbations, this suppression factor is typically around the 10\% level \cite{Qin2025} at low ($z \lesssim 100$) redshifts. No detailed study exists for string-seeded halos, though it is not hard to imagine that the rather exotic nature of the fragmentation procedure could yield large angular momenta, suggesting that $\epsilon_{\rm L} \ll 1$. We leave a more comprehensive study of this to future work.

Neglecting the filament component, we perform the integral and find $N_{\rm DCBH} = \epsilon_{\rm L} \times (0.18, \, 2\times 10^{-3}, \,1.7 \times 10^{-5}) \, {\rm Mpc}^{-3}$ for $G\mu = (10^{-7}, 10^{-8}, 10^{-9})$ respectively. Our consistency with the inferred density \cite{Matthee2023, Maiolino2023} of little red dots detected by JWST ($N_{\rm LRD} \simeq  10^{-5} \, {\rm Mpc}^{-3}$) is thus highly dependent on the exact value of $\epsilon_{\rm L}$.

\section{Discussion and conclusions} \label{sec:discussion}
In this work, we have computed the comoving halo mass functions expected from a distribution of cosmic string loops. These mass functions can be used in a wide variety of future calculations, though here we have placed a particular emphasis on determining the abundance of direct collapse black holes one could expect given a string tension $G\mu$ and velocity distribution function $f(v)$. We have found that reasonable values of these parameters are capable of reproducing the expected abundance of little red dots, though a proper quantitative comparison requires a robust determination of $\epsilon_{\rm L}$ for string-seeded halos.  While this improves on previous work (in particular \cite{Shlaer2012,Jiao2023, Jiao2023b}), there remain many fruitful avenues ripe for follow-up studies. 

First and perhaps most obvious is to critically examine the parameter space in light of constraints on small scale power. As mentioned earlier, string loops do not produce significant spectral distortions \cite{Tashiro2012}, so they are exempt from the current COBE/FIRAS bounds\footnote{Interestingly, if a next generation spectral distortion instrument such as PIXIE or FOSSIL were to confirm that the primordial power spectrum remains nearly scale-invariant on scales $1 \, {\rm Mpc}^{-1} \lesssim k \lesssim 10^4 \, {\rm Mpc}^{-1}$, one could still source direct collapse black holes at high redshift via string loops.} \cite{Chluba2012,Cyr2023b}. An abundance of star forming halos at high redshift can, however, lead to early reionization. Constraints derived by Shlaer \textit{et al.} \cite{Shlaer2012} indicate that tensions with $G\mu \gtrsim 10^{-7}$ reionize the Univere prematurely, under the assumption that $f(v) = \delta(v_{\rm form} - 0.3)$. It would be useful to update these constraints using more modern tools such as \texttt{DarkHistory} \cite{Liu2019, Liu2023, Liu2023b} or \texttt{DM21cm} \cite{Sun2023}.

In addition to these early reionization limits, understanding how constraints from dynamical heating of ultrafaint dwarfs \cite{Graham2024}, CMB accretion, and the microlensing of Icarus\cite{Bringmann2025} translate onto the string parameter space is a clear target for future work. These constraints are nominally given on enhancements to the primordial power spectrum $P_{\zeta}(k)$, which usually lead to the formation of ultra-compact minihalos (UCMHs). The bounds are not directly applicable to string loops as the growth of a string seeded overdensity is qualitatively different than that of an UCMH, and the loops themselves follow a highly non-gaussian distribution. Nevertheless, with the halo mass functions given in this work, translating these bounds should be possible.

Beyond looking at constraints, the formalism presented above is currently limited by hard cutoffs imposed at $M_{\rm min}$ and $M_{\rm max}$ for a given set of parameters. One could extend $M_{\rm min}$ down to lower masses by considering the expected accretion around a string loop which decays shortly after matter-radiation equality. One subtlety associated with these rapidly decaying loops is the so-called rocket effect \cite{Jain2020}, in which gravitational emission of a loop near the end of its lifetime is highly asymmetric. This asymmetric emission causing the loop to accelerate significantly during its death throes. This effect is unimportant for our work, as string-seeded overdensities tend to become self-sufficient roughly one Hubble time after $z_{\rm eq}$, but would need to be taken into account if one was to lower the $M_{\rm min}$ boundary.

On the other end of the spectrum, one could in principle extend $M_{\rm max}$ to higher values by considering the detailed accretion pattern around a (relatively slowly) oscillating loop with $v \simeq \mathcal{O}(0.1)$. Some first steps towards this have been taken in \cite{Jiao2023}, though it is likely that a full numerical setup is necessary to understand the growth rate around these massive loops. 

Recently, full hydrodynamic simulations were performed \cite{Koehler2024} with the inclusion of stationary string loops in an effort to study enhancements to the population of high redshift galaxies. Interestingly, these authors \cite{Koehler2024} found that string loops can match JWST observations surprisingly well. As we have discussed above, however, only a small fraction of string loops are expected to be roughly stationary with respect to the dark matter. It would therefore be interesting to re-run these simulations with our more general mass function derived in Eq.~\eqref{eq:halo_mass_function}.

From the simulation standpoint, it would also be valuable to understand the precise morphology of filaments and beads sourced by string loops. Although not discussed in great detail above, halos which can undergo direct collapse also require low angular momentum (or an efficient way to dissipate it) \cite{Bromm2002,Koushiappas2003,Lodato2006}. Due to the rather exotic formation procedure described here, we have chosen not to speculate on the angular momentum distribution of such halos, and instead defer a detailed analysis of this to future work.

The string-seeded scenario also predicts filamentary correlations between high redshift galaxies. Provided that a cylindrical overdensity fragments, it is reasonable to expect that if one bead in a cylinder can satisfy the direct collapse conditions, the other beads should also form early black holes and star forming regions. This is a rather unique prediction of such models which will be interesting to investigate further in light of current and future data on large scale structure. 

Cosmic string loops remain a well-motivated extension to the standard model of particle physics, containing a rich set of phenomenology with a minimal set of parameters. If they exist, their gravitational influence is expected to provide distinct, non-gaussian imprints on the large scale structure of the Universe. For reasonable values of the string tension $G\mu$, it appears that a significant fraction of string seeded halos are capable of producing direct collapse black holes at very early times. It is therefore possible that the abundance of high redshift black holes and galaxies detected by e.g. JWST have a cosmic string origin.

\section*{Acknowledgments}
BC would like to thank Dominic Agius and Pierre Auclair for insightful discussions on accretion dynamics and loop velocity distributions, respectively. Additionally, the author is thankful to Robert Brandenberger, Wenzer Qin, and Tanmay Vachaspati for insightful comments on an early version of the manuscript. BC is grateful for support from an NSERC Banting Fellowship, as well as the Simons Foundation (Grant Number 929255).

\vspace{30mm}

\appendix

\section{Stationary loop fudge factor} \label{sec:appendix_a}
A simple expression relating $z_{\rm III}$ and $z_{\rm form}$ is given by
%
\begin{align} \label{eq:find_zIII}
    1+z_{\rm form} = \left(\frac{\alpha}{\sigma} \right)^{1/2} &(1+z_{\rm eq}) \left(\frac{1+z_{\rm III}}{1+z_{\rm eq}}\right)^{1/4} \nonumber \\
    &\times \left(\frac{9 \sigma}{5} G\mu \,\tilde{M}(z_{\rm III} )\right)^{-1/4},
\end{align}
%
where $\tilde{M}(z)$ contains terms that become important at $z\simeq z_{\rm eq}$. It is given explicitly as
%
\begin{align}
    \tilde{M}(z) = 1 - \frac{5}{6} \left(\frac{1+z}{1+z_{\rm eq}}\right) - \frac{1}{6}\left(\frac{1+z}{1+z_{\rm eq}}\right)^{5/2}.  \nonumber
\end{align}
%
For a given formation redshift, one uses Eq.~\eqref{eq:find_zIII} to determine $z_{\rm III}$, which can then be utilized to determine the fudge factor $A$ in Eq.~\eqref{eq:fudge_factor_approx}. The full expression that satisfies the matching at between $M_{\rm nl}^{\rm I/II}$ and $M_{\rm nl}^{\rm III}$ at $z_{\rm III}$ is given by
%
\begin{align}
    A &= \frac{M_{\rm nl}^{\rm III}(z_{\rm III})}{M_{\rm nl}^{\rm I}(z_{\rm III})}, \nonumber\\
    &=\left(\frac{\pi}{2} \right)^{3/2} \left[\frac{1 - \frac{5}{6} \left(\frac{1+z_{\rm III}}{1+z_{\rm eq}}\right) - \frac{1}{6}\left(\frac{1+z_{\rm III}}{1+z_{\rm eq}}\right)^{5/2}}{\log \left(\frac{1+z_{\rm eq}}{1+z_{\rm III}}\right) + \frac{1}{10} - \frac{1}{10}\left(\frac{1+z_{\rm III}}{1+z_{\rm eq}}\right)^{5/2}} \right]^{3/2}.
\end{align}
%
The logarithmic term dominates for late times when $z_{\rm III} \ll z_{\rm eq}$. Note that when $z_{\rm III} = z_{\rm eq}$, the fudge factor is indeterminate. This is an artifact of us neglecting any growth in the radiation era, and doesn't cause any issues in practice as $z_{\rm III} = z_{\rm eq}$ implies we should only be utilizing $M_{\rm nl}^{\rm III}$. In situations where $z_{\rm III} < 0$ (that is, when the point mass approximation isn't valid at any redshift), one should resort back to utilizing $M_{\rm nl}^{\rm I}(z)$ at all times.

\section{Jacobian factors} \label{sec:Jacobian_Factor}
For completeness, we list the Jacobian factors used in computing the halo mass spectrum, Eq.~\eqref{eq:halo_mass_function}. There are four possible accretion scenarios, with three of them being trivial to calculate,
%
\begin{align}
    \frac{\partial L}{\partial M_{\rm nl}^{\rm III}} &= \frac{5}{2 \mu \tilde{M}(z)}\left(\frac{1+z}{1+z_{\rm eq}} \right), \nonumber\\
    \frac{\partial L}{\partial M_{\rm fil}} &= \frac{5}{3 \mu \tilde{M}(z)}\left(\frac{1+z}{1+z_{\rm eq}} \right), \nonumber\\
    \frac{\partial L}{\partial M_{\rm bead, N\geq 1}} &= \frac{4}{3(4\pi)^{4/3}}G \left( \frac{G}{t_{\rm eq}}\right)^{1/3} \frac{v_{\rm form}^2}{\alpha (G\mu)^2}, \nonumber \\
    &\,\,\,\,\,\,\,\, \times \left( \frac{\tilde{L}}{\tilde{M}}\right)^{4/3} M^{1/3}_{\rm bead, N\geq 1} \left(\frac{1+z}{1+z_{\rm eq}} \right)^2. \nonumber
\end{align}
%
Interestingly, the growth rate for the stationary case in Regions I and II ($M_{\rm nl}^{\rm I/II}$) does not have a dependence on $L$ and thus the Jacobian appears infinite. This is ultimately a side effect of the $\arcsin(aq/R) \simeq aq/R$ approximation that was used to derive the Region I growth. The full solution does contain an $R$ dependence (see the appendix of Hao \textit{et al.} \cite{Jiao2023}), and can be used to safely compute this Jacobian. However, as shown above, the fraction of loops in a realistic velocity distribution undergoing stationary Region I or II growth is minuscule. It is therefore a reasonable approximation to simply neglect the contribution to the halo mass function from that subdominant population of loops. 

It may be useful for the reader to note that the formation redshift and length of a loop can be approximately related by
%
\begin{align}
    L \simeq \alpha t_{\rm eq} \left(\frac{1+z_{\rm eq}}{1+z_{\rm form}} \right)^2.
\end{align}
%
From this expression one can further relate $z_{\rm form}$ to the halo mass through the expressions given above.

\section{Minimum and maximum halo masses} \label{sec:Mass_range}
The calculations presented in this work make various assumptions, which in turn limit the range of applicable masses in the halo and black hole mass functions. Here, we derive the precise range of validity given these assumptions. The smallest loops we consider are the ones who exist for roughly one Hubble time after matter radiation equality. This is because it takes roughly a Hubble time for the halo or filament to accumulate a total mass greater than the mass of the loop, namely when $M_{\rm nl} \geq \mu L$. Thus, our smallest loops have a length set by the requirement that they decay at $z_{\rm dec} \simeq  z_{\rm eq}/2$, yielding 
%
\begin{align}
    L_{\rm min} = \Gamma_{\rm g} G\mu \,t_{\rm eq} \left(\frac{1+z_{\rm eq}}{1+z_{\rm eq}/2} \right)^{3/2}.
\end{align}
%
On the opposite end of the spectrum, we consider our calculations to be robust for loops which formed roughly one Hubble time before matter radiation equality, namely $z_{\rm form} \geq 2 z_{\rm eq}$. Loops formed at this redshift will be small enough at matter radiation equality such that oscillations about their centre of mass will be rapid enough that the density profile of the loop will appear smeared out from the perspective of infalling mass shells. This give us
%
\begin{align}
    L_{\rm max} = \alpha t_{\rm eq} \left(\frac{1+z_{\rm eq}}{1+2z_{\rm eq}} \right)^2.
\end{align}
%
Larger loops, especially those formed at $z_{\rm form} \simeq z_{\rm eq}$ will in principle grow to even larger halos, but require a more detailed study which we defer to future work.

From here, we can determine the minimum and maximum halo masses in the unbroken and fragmented filament scenarios. For the filament case, we simply find
%
\begin{align}
    M_{\rm fil}^{\rm min} &\simeq 3.7 \times 10^{3} \, M_{\odot} \left( \frac{G\mu}{10^{-8}}\right)^2 \left( \frac{1+z_{\rm eq}}{1+z}\right) \tilde{M}(z), \nonumber \\
    M_{\rm fil}^{\rm max} &\simeq 6.5 \times 10^{7} \, M_{\odot} \left( \frac{G\mu}{10^{-8}}\right) \left( \frac{1+z_{\rm eq}}{1+z}\right) \tilde{M}(z). \nonumber
\end{align}
%
Note that the total range is set by $G\mu$, with smaller values leading to a larger range of masses for which our computation is valid. This makes sense, as a smaller string tension prolongs the lifetime of any given loop. Similarly, when the filament fragments into beads, the valid range of bead masses is given by
%
\begin{align}
    M_{\rm bead}^{\rm min} &\simeq 2.4 \times 10^2 \, M_{\odot} \left(\frac{0.3}{v_{\rm form}} \right)^{3/2} \left( \frac{G\mu}{10^{-8}}\right)^{9/4} \nonumber\\
    &\hspace{30mm} \times \left(\frac{1+z_{\rm eq}}{1+z} \right)^{3/2} \frac{\tilde{M}(z)}{\tilde{L}(z)},\nonumber\\
    M_{\rm bead}^{\rm max} &\simeq 3.7 \times 10^5 \, M_{\odot} \left(\frac{0.3}{v_{\rm form}} \right)^{3/2} \left( \frac{G\mu}{10^{-8}}\right)^{3/2} \nonumber\\
    &\hspace{30mm} \times \left(\frac{1+z_{\rm eq}}{1+z} \right)^{3/2} \frac{\tilde{M}(z)}{\tilde{L}(z)}.\nonumber
\end{align}
%
While this looks slightly more complicated as there is a $v_{\rm form}$ dependence, we remind the reader that once $v_{\rm form} \leq v_{\rm form}^{\rm frag}$, the filament masses should be used.

\clearpage
\bibliographystyle{apsrev4-1}
\bibliography{Lit}

\end{document}